 \documentclass[final,3p,times,twocolumn]{elsarticle}

\usepackage{amssymb}

\usepackage{lineno}

\usepackage{unites2e}

\newcommand{\hgc}{HESS~J1745-290}
\newcommand{\astar}{Sgr~A$^*$}
\newcommand{\aeast}{Sgr~A~East}
\newcommand{\pwn}{G359.95-0.04}

\journal{Astroparticle Physics}

\begin{document}

\begin{frontmatter}


\ead{jholder@physics.udel.edu }


\title{TeV Gamma-ray Astronomy: A Summary}


\author[1]{J. Holder}

\address[1]{Department of Physics and Astronomy and the Bartol Research Institute, University of Delaware, Newark, DE 19716, USA, ~~~~~~~~~~~~~~~~~~~~~~~~~~~~~~~~~~~~Tel: +(1) 302 831 2545, Fax: +(1) 302 831 1637}

\begin{abstract}
The field of TeV gamma-ray astronomy has produced many exciting
results over the last decade. Both the source catalogue, and the range
of astrophysical questions which can be addressed, continue to expand.
This article presents a topical review of the field, with a focus on
the observational results of the imaging atmospheric Cherenkov
telescope arrays. The results encompass pulsars and their nebulae,
supernova remnants, gamma-ray binary systems, star forming regions and
starburst and active galaxies.

\end{abstract}

\begin{keyword}
Gamma-ray Astronomy \sep
High-energy astrophysics


\end{keyword}

\end{frontmatter}


\section{Introduction}
\label{intro}

Teraelectronvolt (TeV) astronomy concerns the study of astrophysical
sources of gamma-ray photons, with energies in the range between
$\sim30\U{GeV}$ and $\sim30\U{TeV}$. The TeV range is one of the most
recent windows of the electromagnetic spectrum to be opened for study,
beginning with the identification of the first source, the Crab
Nebula, in 1989 \cite{1989ApJ...342..379W}.  The results have been
impressive, with significant advances in instrumentation leading to
the detection of well over 100 sources over the last decade.  The
goals of TeV astronomy are wide-ranging, but can broadly be described
as the study of sites of relativistic particle acceleration in the
Universe, both hadronic and leptonic. This encompasses a huge range of
size scales and energetics, from the interactions of galaxy clusters,
to the magnetospheres of individual pulsars.

Numerous recent reviews of TeV gamma-ray astronomy have been written
(e.g. \cite{2009ARA&A..47..523H, 2008RPPh...71i6901A,
2008JPhG...35c3201C}), but the field is rapidly evolving. For example,
since the extensive review by Hinton \& Hofmann in 2009
\cite{2009ARA&A..47..523H}, the number of sources has grown from
around 80 to more than 120, and a number of new source classes have been
identified. This paper therefore aims to provide an update, and to
supplement the existing reviews with a summary of the current
observational status. The primary focus of the review is on the
results from imaging atmospheric Cherenkov telescopes, with a
relatively brief discussion of the air shower particle detector
experiments. We use the definitions outlined by Aharonian
\cite{2004vhec.book.....A} as follows: ``high energy'', or ``GeV'',
astronomy refers to the energy range from $30\U{MeV}$ to $30\U{GeV}$,
while ``very high energy'', or ``TeV'', astronomy, refers to the range
from $30\U{GeV}$ to $30\U{TeV}$.

\section{A Brief History}

The development of ground-based gamma-ray astronomy is closely linked
to the study of cosmic rays and cosmic ray air showers. The idea of
searching for astrophysical gamma-ray sources at $\sim100\U{MeV}$
energies was first proposed by Morrison in 1958 \cite{Morrison58}. A
prediction of a very high \textit{TeV} gamma-ray flux from various
sources, including the Crab, which might be detectable with an air
shower particle array at high altitude, was made by Cocconi in 1959
\cite{Cocconi59}. Cherenkov radiation associated with large cosmic ray
air showers was first detected by Galbraith \& Jelley in 1953
\cite{1953Natur.171..349G}, and the possibility of using this
phenomenon to study gamma-ray initiated showers led to the development
of a number of dedicated facilities in the 1960s. This effort was
boosted by the apparent detection of a gamma-ray signal from the
black-hole binary Cygnus X-3 by both particle air shower arrays and
atmospheric Cherenkov detectors. With hindsight, this detection was
likely spurious, as were numerous unsubstantiated claims throughout
the 1970s, 1980s and into the 1990s.

The field reached a firm experimental footing with the development of
the imaging technique, which provides a method of effectively
discriminating between gamma-ray initiated showers and the background
of cosmic ray showers, based on the morphology of their Cherenkov
images, and guided by the results of Monte Carlo simulations
\cite{1977ESASP.124..279W, 1985ICRC....7..231H}. This technique was
applied by the Whipple collaboration to detect steady gamma-ray
emission from the Crab Nebula using a 10m reflector with a 37-element
photomultiplier tube camera in 1989 \cite{1989ApJ...342..379W}. A
number of imaging atmospheric Cherenkov telescopes (IACTs) were
subsequently developed around the world (including Durham, CANGAROO,
Telescope Array, Crimean Astrophysical Observatory, SHALON, TACTIC),
with the northern hemisphere instruments (Whipple, HEGRA and CAT)
leading the field. The 1990s saw two particularly important
developments: the detection of the first extragalactic sources by the
Whipple Collaboration, starting with the nearby blazars Markarian~421
\cite{1992Natur.358..477P} and Markarian~501
\cite{1996ApJ...456L..83Q}, and the application of the stereo imaging
technique by the HEGRA array \cite{1997APh.....8....1D}. HEGRA
consisted of 5 telescopes of modest aperture ($<10\UU{m}{2}$), and
demonstrated that the combination of Cherenkov image information from
multiple telescopes located within the same Cherenkov light pool could
dramatically improve the sensitivity of the technique.

Despite this progress, the relative scarcity of bright TeV gamma-ray
sources ($<10$ were identified by 2000) highlighted the necessity for
improved instrumentation. Cherenkov wavefront samplers such as CELESTE
and STACEE attempted to probe to lower energies, and hence higher
gamma-ray fluxes and larger distances, using converted solar farms;
however, the difficulty of discriminating gamma-rays from the cosmic
ray background using this technique limited its
effectiveness. Successful gamma-ray observations using a particle
detector were made by the Milagro experiment, which ran from 2000 to
2008, providing a survey of the northern sky with modest
sensitivity. Starting with the commissioning of H.E.S.S. in 2003, the
new generation of IACTs - H.E.S.S., MAGIC and VERITAS - have provided
the required order of magnitude improvement in sensitivity, and firmly
established gamma-ray studies as an important astronomical discipline.

\section{Current Status and Instrumentation}
\subsection{Imaging Atmospheric Cherenkov Telescopes}

The Cherenkov emission from an air shower forms a column of blue light
in the sky, with the maximum emission occuring around $10\U{km}$ above
sea level at TeV energies. Cherenkov telescopes used to record these
images are essentially rather simple devices, consisting of a large,
segmented optical flux collector used to focus the Cherenkov light
onto an array of fast photo-detectors. The optical specifications are
not terribly strict; an optical point spread function width of typically
$\sim0.05^{\circ}$ is adequate. The photo-detector array (usually
$<1000$ photomultiplier tubes) comprises a crude camera, covering a
few degrees on the sky. A Cherenkov flash triggers read-out of the
photo-detectors, with a read-out window defined by the timescale of
the arrival of the Cherenkov photons ($\sim10\U{ns}$).

There are currently three major imaging atmospheric Cherenkov
telescope systems in operation. H.E.S.S., located in the Khomas
Highland of Namibia ($-23^{\circ}$N, $-16^{\circ}$W, altitude
$1800\U{m}$), consists of four telescopes arranged on a square with
120\U{m} side length. Each telescope has a mirror area of
$107\UU{m}{2}$ and is equipped with a 960 pixel camera covering a
$5^{\circ}$ field of view. VERITAS, at the Fred Lawrence Whipple
Observatory in southern Arizona ($32^{\circ}$N, $111^{\circ}$W,
altitude $1275\U{m}$) has similar characteristics, with 4 telescopes
of $107\UU{m}{2}$ area and 499-pixel cameras, covering $3.5^\circ$.
MAGIC ($28^{\circ}$N, $17^{\circ}$W, altitude $2225\U{m}$) originally
consisted of a single, very large reflector ($236\UU{m}{2}$) on the
Canary island of La Palma, with a $3.5^{\circ}$ camera. In 2009, a
second telescope with the same mirror area was installed at a distance
of $85\U{m}$ from the first.

\begin{figure}[!t]
  \centering
  \includegraphics[width=\columnwidth]{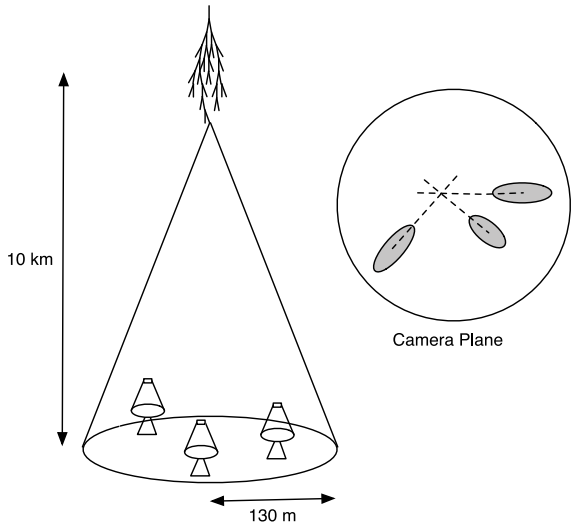}
  \caption{A schematic illustration of an atmospheric Cherenkov
telescope array. The primary particle initiates an air shower,
resulting in a cone of Cherenkov radiation. Telescopes within the
Cherenkov light pool record elliptical images; the intersection of the
long axes of these images indicates the arrival direction of the
primary, and hence the location of a $\gamma$-ray source in the sky.}
  \label{ACT}
 \end{figure}

Each of these facilities work in a similar fashion. Cherenkov images
of air showers are recorded at a rate of a few hundred Hz, and
analyzed offline. The overwhelming majority of these images are due to
cosmic ray initiated air showers. Gamma-ray showers can be
discriminated from this background based on the image shape and
orientation. Gamma-ray images result from purely electromagnetic
cascades and appear as narrow, elongated ellipses in the camera
plane. The long axis of the ellipse corresponds to the vertical
extension of the air shower, and points back towards the source
position in the field of view. If multiple telescopes are used to view
the same shower, the source position is simply the intersection point
of the various image axes (illustrated schematically in
Figure~\ref{ACT}). Cosmic-ray primaries produce secondaries with large
transverse momenta, which initiate sub-showers. The Cherenkov images
of cosmic-ray initiated air showers are consequently wider then those
with $\gamma$-ray primaries, and form an irregular shape, as opposed
to a smooth ellipse. In addition, since the original charged particle
has been deflected by galactic magnetic fields before reaching the
Earth, cosmic-ray images have no preferred orientation in the camera.

Cherenkov light reaching the ground from air showers peaks at
optical/UV wavelengths, and so IACTs operate only under clear, dark
skies. Both MAGIC and VERITAS have demonstrated that useful
observations can be made when the moon is visible above the horizon,
but the typical duty cycle of these instruments is still limited to
$\sim1200\U{hours}$ per year ($<15\%$). Given the small field of view
of IACTs, regions of the sky containing one or more source candidates
are usually targeted for observations. Surveys can only be
accomplished slowly, by tiling regions of the sky with overlapping
fields-of-view. The sensitivity of the current generation of IACTs is
sufficient to detect the Crab Nebula in under a minute, and a source
with 1\% of the Crab flux ($\sim2\times10^{-13}\UU{m}{-2}\UU{s}{-1}$
above $1\U{TeV}$) in $\sim25\U{hours}$. The angular and energy
resolution of the technique are energy-dependent, with typical values
of $<0.1^{\circ}$ and $<15\%$ per photon, respectively, at $1\U{TeV}$.

\begin{figure*}[!t]
  \centering \includegraphics[width=\textwidth]{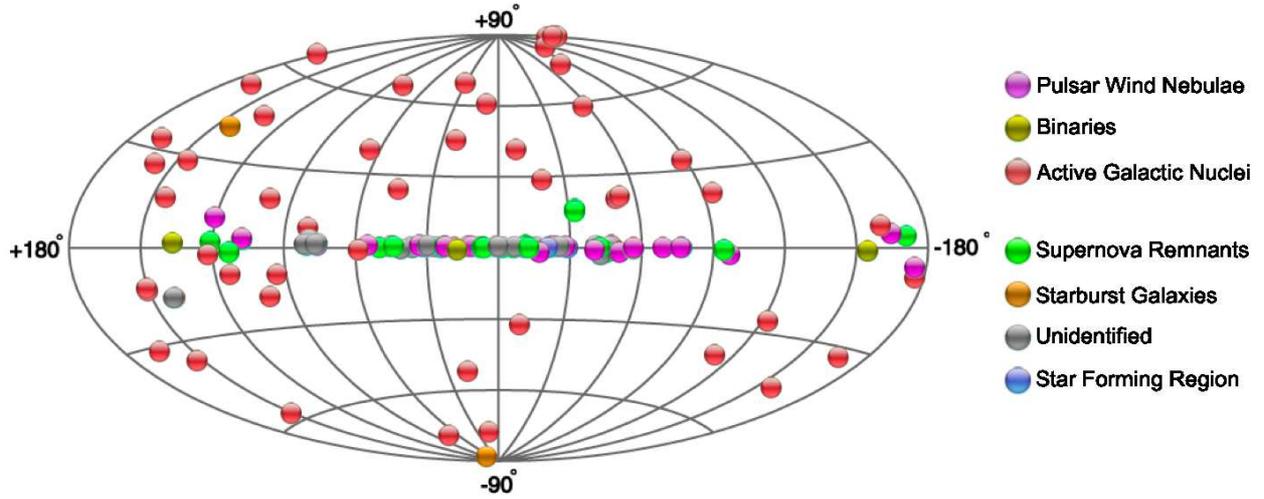}
  \caption{A map of the catalog of localized sources of TeV gamma-ray
  emission in Galactic coordinates as of November 2011, provided by
  the online catalog \textit{TeVCat}
  \cite{2008ICRC....3.1341W} 
  }
  \label{TeVCat}
 \end{figure*}

The catalog of TeV sources grew rapidly with the commissioning of
H.E.S.S. in the southern hemisphere, which provided the first high
sensitivity observations of the densely populated inner Galaxy. It has
continued to expand in recent years as MAGIC and VERITAS have come
online, and now numbers 130 sources, as listed in the online catalog
\textit{TeVCat} \cite{2008ICRC....3.1341W} \footnote{We note that a
small fraction of these 130 sources may be duplicate detections of the
same object, while other detections remain contentious or
unconfirmed. Full details are available in the individual source
annotations in \textit{TeVCat}, available at
{http://tevcat.uchicago.edu}}. Figure~\ref{TeVCat} shows the locations
of these sources in Galactic coordinates.

\subsection{Particle Detectors}

The direct detection of air shower particles using arrays of particle
detectors at ground level offers some important advantages over the
atmospheric Cherenkov technique. In particular, observations can be
made continuously, both day and night, and over the entire viewable
sky (a field-of-view of $>1\U{steradian}$). These advantages are
offset, however, by the rather low sensitivity to point sources, which
is primarily due to the difficulty of rejecting the substantial
background of cosmic ray initiated air showers. The angular and energy
resolution of these detectors are also significantly worse than
IACTs. Early claims for emission from binary systems using sparse
particle detector arrays were not confirmed by later, more sensitive
instruments (e.g. \cite{1993ApJ...417..742M, 2002A&A...390...39A}),
indicating the need for a new approach to this problem. Milagro, which
operated between 2000 and 2008 in northern New Mexico ($36^{\circ}$N,
$107^{\circ}$W), was the first successful attempt at this.

\begin{figure}[!t]
  \centering \includegraphics[width=\columnwidth]{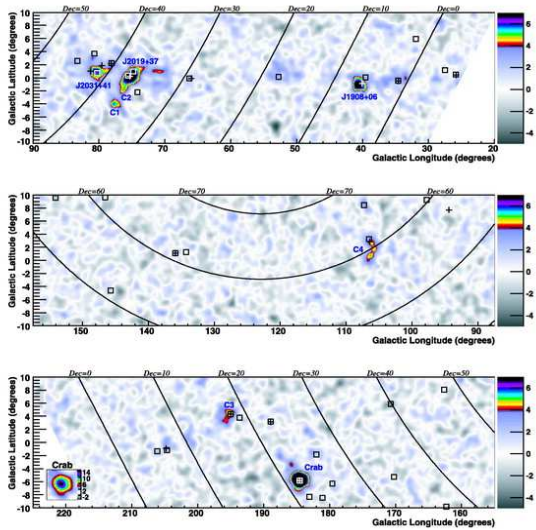}
  \caption{The Milagro survey of the Galactic plane. The
  \textit{z}-axis is the pre-trials statistical significance, with a
  fixed maximum of $7\sigma$. Figure from Abdo et
  al. \cite{2007ApJ...664L..91A}.  }
  \label{Milagro}
 \end{figure}

The Milagro detector consisted of a large water reservoir
($60\times80\U{m}$) at an altitude of $2630\U{m}$, covered with a
light-tight barrier, and instrumented with PMTs. The central reservoir
provided high-resolution sampling of air shower particles over a
relatively small area (compared to the air shower footprint). In 2004
an array of 175 small outrigger tanks were added, irregularly spread
over an area of $200\times200\U{m}$ around the central reservoir. This
configuration, coupled with the development of analysis techniques for
cosmic ray background discrimination, provided sufficient sensitivity
for the first comprehensive survey of the northern TeV sky. The
results showed strong detections of the bright, known TeV sources
Markarian 421 and the Crab Nebula, along with the detection of three
extended sources in the Galactic plane, each with integrated fluxes
comparable to the Crab Nebula at $20\U{TeV}$. A few less significant
source candidates were also identified in the plane, and a reanalysis
following the launch of \textit{Fermi}-LAT also showed fourteen $3\sigma$
excesses co-located with bright Galactic LAT sources
\cite{2009ApJ...700L.127A}. The Milagro results for the region around
the Galactic plane are shown in Figure~\ref{Milagro}.

Milagro ceased operation in 2008; however, two large particle detector
arrays remain in operation at very high altitude in Tibet. ARGO-YBJ is
located at $4300\U{m}$ in Yangbajing, and consists of a single layer
of resistive plate chambers completely covering an area of
$110\times100\U{m}$. The results of 1265 days of observations were
recently presented \cite{2011arXiv1110.1809C}, showing $>5\sigma$
detections of the Crab Nebula, Markarian~421 and two Milagro
sources. The fact that one of the brightest Milagro sources,
MGRO~J2019+37, is not detected in these observations presents
something of a mystery. The Tibet AS$\gamma$ air shower array, also at
Yangbajing, consists of $\sim750$ closely-spaced scintillation
detectors covering an area of $36900\UU{m}{2}$, and has demonstrated
that this technique is also practical for the detection of bright TeV
sources \cite{2009ApJ...692...61A}.


\section{Extragalactic TeV Sources}
\subsection{Blazars}

Approximately 1\% of all galaxies host an active nucleus; a central
compact region with much higher than normal luminosity. Around 10\% of
these Active Galactic Nuclei (AGN) exhibit relativistic jets, powered
by accretion onto a supermassive black hole. Many of the observational
characteristics of AGN can be attributed to the geometry of the
system; in particular, the orientation of the jets with respect to the
observer. Blazars, which host a jet oriented at an acute angle to the
line of sight, are of particular interest for gamma-ray astronomy, as
the emission from these objects is dominated by relativistic beaming
effects, which dramatically boost the observed photon energies and
luminosity.

 \begin{figure*}[!t]
  \centering 
  \includegraphics[height=1.6in]{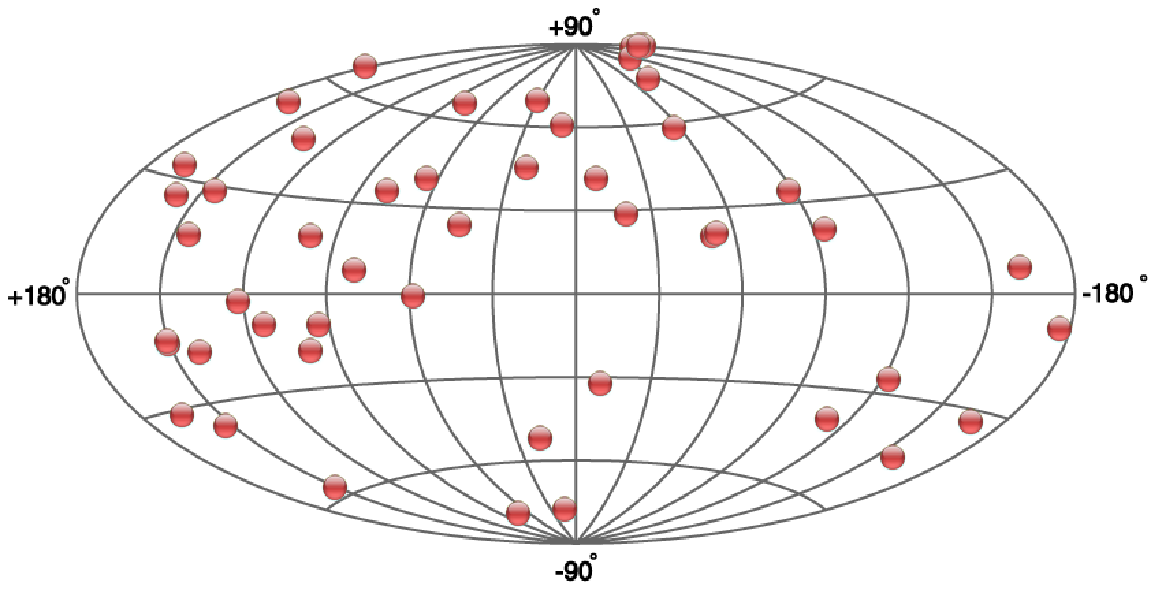}\includegraphics[height=1.6in]{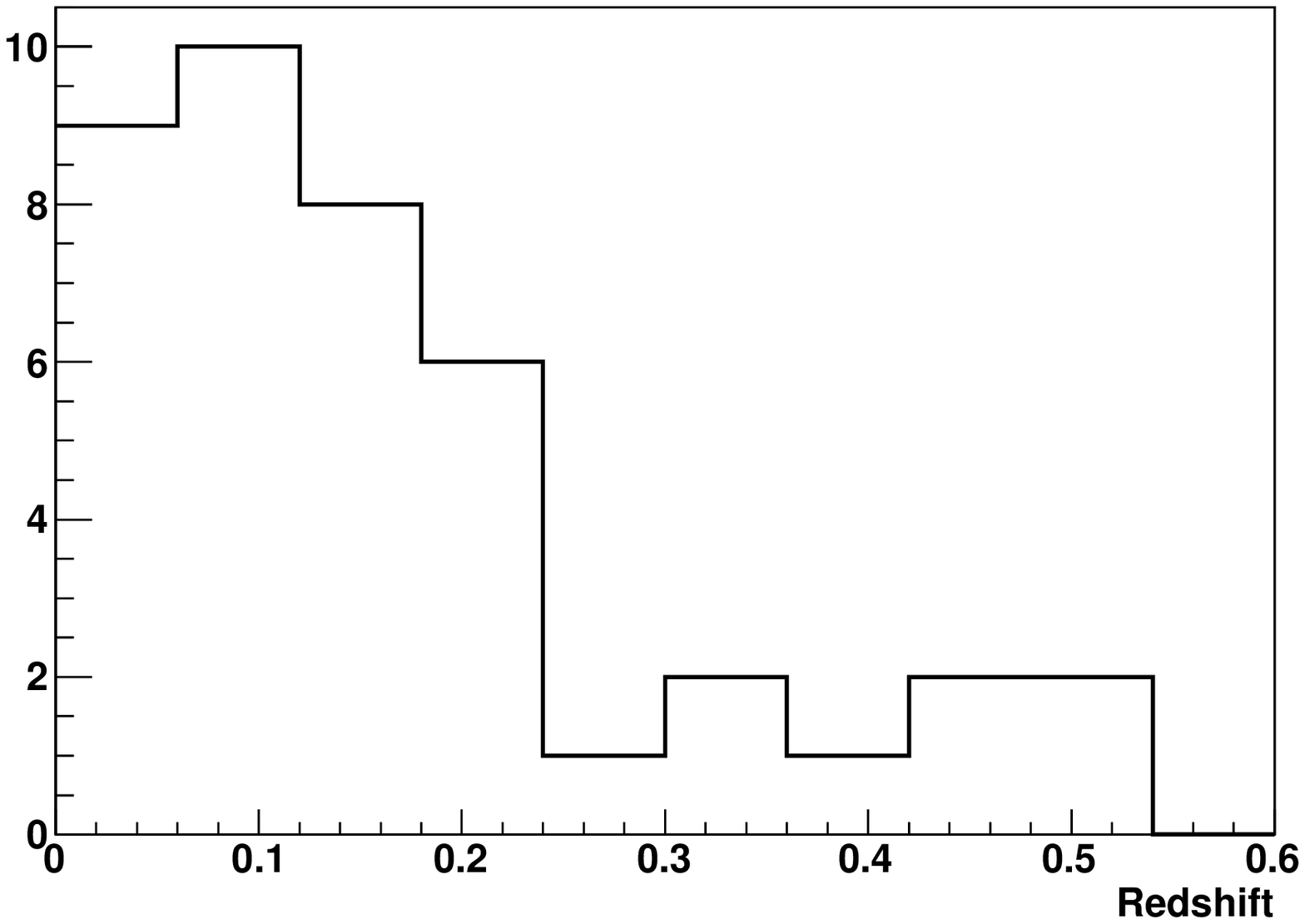}
  \caption{The location of the 49 AGN (BL Lac, FSRQs and radio
galaxies) with detected TeV emission (from \textit{TeVCat}
\cite{2008ICRC....3.1341W}). On the left is the spatial distribution,
in Galactic coordinates; The right plot shows the redshift
distribution, for 41 of the objects with redshifts listed in the
literature.  }
  \label{map_redshifts}
 \end{figure*}

\begin{figure*}[!t]
  \centering 
  \includegraphics[width=0.8\textwidth]{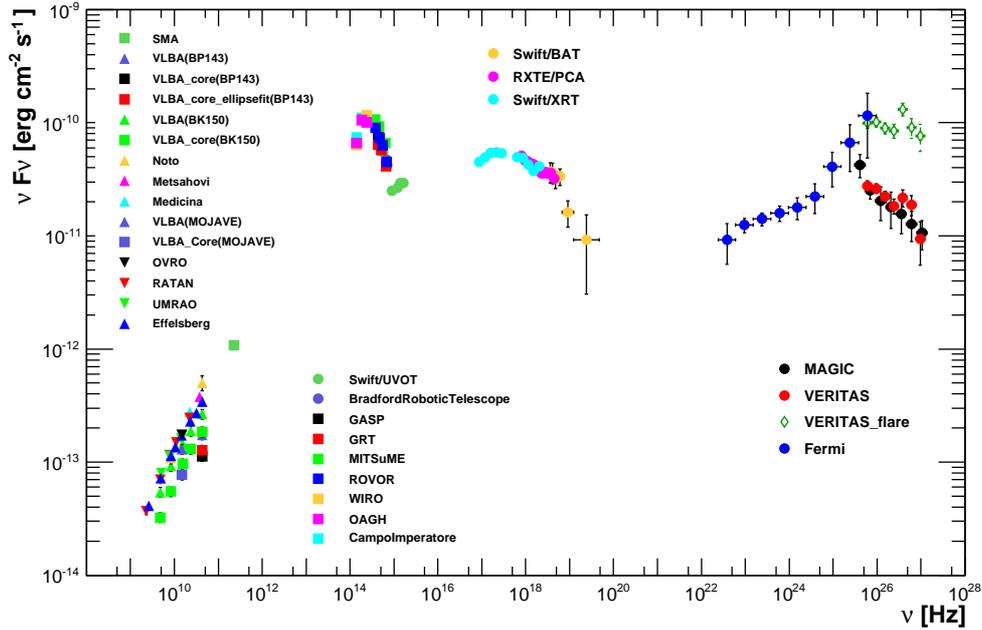}
  \caption{ Extensive multifrequency measurements showing the spectral
  energy distribution of Markarian 501 for observations in
  2009. Emission from the host galaxy is clearly visible at
  infrared/optical frequencies. The VERITAS data are divided to show
  both the average spectrum (red circles), and the spectrum during a
  3-day flare (green triangles). Figure from Abdo et
  al. \cite{2011ApJ...727..129A} }
  \label{SED}
 \end{figure*}

The first extragalactic source discovered at TeV energies was
Markarian 421 \cite{1992Natur.358..477P}, a blazar of the BL Lacertae
sub-class. The extragalactic TeV catalog now comprises $\sim50$
objects, and continues to increase steadily
(Figure~\ref{map_redshifts}). Blazar SEDs show a double-peaked
structure in a $\nu F_{\nu}$ representation of their spectral energy
distribution (SED), with the lower frequency peak usually attributed
to synchrotron emission of energetic electrons, and the higher
frequency peak to inverse Compton. BL Lac objects are further
classified as low-, intermediate- or high-frequency peaked, according
to the location of the peak of their synchrotron emission.  The
majority ($\sim80\%$) of the known TeV blazars are high-frequency
peaked objects, in part because of inherent biases in the target
selection: initially, objects were chosen based primarily upon their
radio and X-ray spectral properties
(e.g. \cite{2002A&A...384...56C}). More recently, the TeV
observatories have expanded their selection criteria, using additional
guidance in the form of \textit{Fermi}-LAT results, and
multi-wavelength observation triggers. This has broadened the catalog
to include examples of intermediate- and low-frequency peaked
objects. The overall data quality has also improved markedly since the
launch of \textit{Fermi}: Figure~\ref{SED} shows a recent compilation
of spectral measurements for the bright TeV blazar, Markarian 501
\cite{2011ApJ...727..129A}, taken during an extensive multiwavelength
campaign in 2009.

The mechanisms which drive the high energy emission from blazars
remain poorly understood, and a full discussion is beyond the scope of
this review.  Briefly; in leptonic scenarios, a population of
electrons is accelerated to TeV energies, typically through Fermi
acceleration by shocks in the AGN jet. These electrons then cool by
radiating X-ray synchrotron photons. TeV emission results from inverse
Compton interactions of the electrons with either their self-generated
synchrotron photons, or an external photon field. The strong
correlation between X-ray and TeV emission which is often observed
provides evidence for a common origin such as this, although
counter-examples do exist \cite{2004ApJ...601..151K}. Another class of
models has hadrons as the primary particle population, which can then
produce TeV gamma-rays through subsequent interactions with target
matter or photon fields. Hadronic models are less favoured, typically,
in part because the cooling times for the relevant processes are long,
making rapid variability difficult to explain. One exception to this
is the case of proton synchrotron emission, which may provide a
plausible alternative, in which the emission results from extremely
high energy protons in highly magnetized ($B\sim100\U{G}$), compact
regions of the jet \cite{2000NewA....5..377A}. 

\begin{figure}[!t]
  \centering \includegraphics[width=\columnwidth]{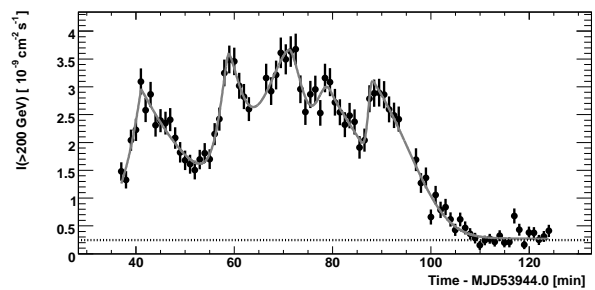}
  \caption{ Integrated flux ($>200\U{GeV}$) versus time for
  H.E.S.S. observations of PKS~2155-304 on MJD~53944 (28 July,
  2006). The data are binned in 1-minute intervals, and the horizontal
  dashed line shows the steady flux from the Crab Nebula for
  comparison. Figure from \cite{2007ApJ...664L..71A}.}
  \label{variability}
 \end{figure}

Many of the AGN detected at TeV energies exhibit extreme
variability. The timescales can range from years to minutes, and the
observed flux can change by more than an order of
magnitude. Figure~\ref{variability} shows the H.E.S.S. lightcurve from
July 2006 for one of the most extreme examples, the BL Lac object
PKS~2155-304 \cite{2007ApJ...664L..91A, 2010A&A...520A..83H}. Such
rapid variability can be used to place constraints on the size of the
emission region, which depend upon the Doppler factor, $\delta$.
$\delta$ itself is constrained by the requirement that the emission
region should be transparent to gamma-rays
(e.g. \cite{1996Natur.383..319G}). Extremely rapid TeV gamma-ray
variability of distant blazars can also be used to place limits on the
energy-dependent violation of Lorentz invariance
\cite{1999PhRvL..83.2108B, 2011APh....34..738H}, as predicted in some
models of quantum gravity.

Related to the BL Lacertae objects are Flat Spectrum Radio Quasars
(FSRQs). These are characterized primarily by their intense UV
emission, associated with an accretion disk, strong broad emission
lines in the optical spectrum, and infra-red emission associated with
a dusty torus. FSRQs are similar to low-frequency peaked BL Lacs, in
that the X-ray emission is dominated by the inverse Compton peak of
the SED. Despite this, inverse Compton emission can extend up to TeV
energies, particularly during intense flaring episodes, and three
FSRQs have recently been detected by IACTs (PKS~1222+21
\cite{2011ApJ...730L...8A} , PKS 1510-089 \cite{2010HEAD...11.2706W}
and 3C279 \cite{2011A&A...530A...4A}). 

A unique additional case is the TeV detection of IC~310, a
``head-tail'' radio galaxy in the Perseus cluster, possibly hosting a
low-luminosity BL Lac nucleus \cite{1999ApJ...516..145R}. Head-tail
radio galaxies display a distinctive radio morphology, consisting of a
bright ``head'' and a fainter ``tail'', which is believed to be the
result of their rapid motion with respect to the intracluster
medium. The source was originally identified as a VHE emitter in an
analysis of the highest energy ($>30\U{GeV}$) Fermi photons by Neronov
et al. \cite{2010A&A...519L...6N}, and then subsequently detected
from the ground by MAGIC \cite{2010ApJ...723L.207A}. Neronov et
al. considered the intriguing possibility that the emission might
originate with particles accelerated at the bow shock formed by the
interaction between the relativistic outflow from the galaxy and the
intracluster medium. This scenario is ruled out by the detection of
variability in the TeV flux by MAGIC, and the more familiar BL Lac
mechanisms are now favoured.

Figure~\ref{map_redshifts} also shows the distribution of measured
redshifts for 41 TeV AGN. Some of these measurements are rather
uncertain, since BL Lac optical spectra, by definition, do not contain
strong emission lines. The most distant object detected is 3C279
\cite{2011A&A...530A...4A}, with a relatively modest redshift of
$z=0.5362$. The population is truncated at large distances due to the
absorption of TeV gamma-rays by electron-positron pair production with
the low energy photons of the extragalactic background light
(EBL). This effect is energy dependent, and can strongly modify the
observed VHE spectra of extragalactic sources. While this limits the
observation of more distant TeV sources, it also provides a mechanism
by which to infer the intensity of the EBL, using reasonable
assumptions about the intrinsic TeV spectra at the source
\cite{1992ApJ...390L..49S}. The EBL provides a calorimetric measure of
the complete history of star and galaxy formation in the Universe, but
is difficult to measure directly, due to the presence of bright local
foreground sources of emission, such as zodiacal light. Presently, all
of the TeV blazar measurements are consistent with a relatively low
level of EBL, with the constraints derived from VHE measurements now
approaching the lower limits derived from galaxy counts
\cite{2006Natur.440.1018A, 2008Sci...320.1752M, 2011ApJ...733...77O}.

TeV blazar observations have also been suggested as probes of other
physical phenomena, such as the acceleration and propagation of
ultra-high energy cosmic rays \cite{1994ApJ...423L...5A,
2007Ap&SS.309..465G}, or, more speculatively, the production of
axion-like particles \cite{2007PhRvL..99w1102H,
2011JCAP...11..020D}. Various authors have also discussed the
possibility that TeV observations may be used to measure or constrain
the strength of the intergalactic magnetic field (IGMF)
(e.g. \cite{1994ApJ...423L...5A, 2011MNRAS.414.3566T}). Temporal,
spectral and spatial signatures of the IGMF are all possible; however,
the fact that blazars are intrinsically variable gamma-ray sources
limits the power of this technique. Accounting for this, Dermer et
al. derive a lower limit of $B_{IGMF}\geq10^{-18}\U{G}$
\cite{2011ApJ...733L..21D}.

\subsection{Radio Galaxies}
As described above, the TeV fluxes from blazars are dramatically
enhanced by the effects of Doppler boosting. Nearby radio galaxies, in
which the jet is not directly oriented towards the line-of-sight,
provide an alternative method by which to investigate the particle
acceleration and gamma-ray emission from relativistic outflows in
AGN. The advantage of studying such objects lies in the fact that the
jets can be resolved from radio to X-ray wavelengths, allowing the
possibility of correlating the gamma-ray emission state with observed
changes in the jet structure. Three radio galaxies have been
identified as TeV emitters: M~87, Centaurus~A and NGC~1275.

\begin{figure}[!t]
  \centering \includegraphics[width=\columnwidth]{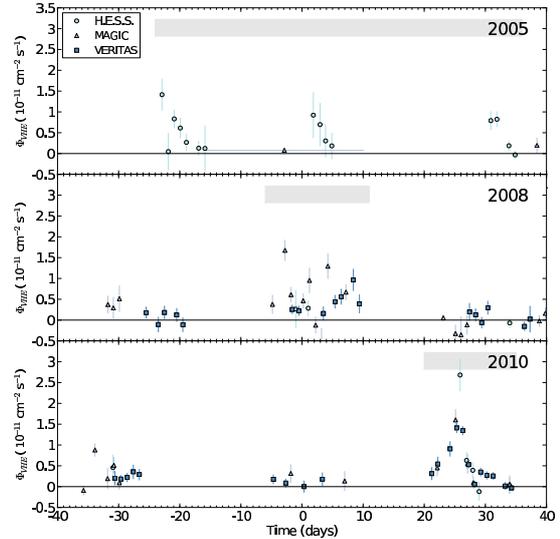}
  \caption{ VHE light curves of the three flares from M87 observed in
  2005, 2008 and 2010, showing integral fluxes above an energy of
  $350\U{GeV}$. Figure from \cite{2012ApJ...746..151A}.}
  \label{M87}
 \end{figure}

M~87 is the most well studied of these, and was first reported as a
gamma-ray source by the HEGRA collaboration
\cite{2003A&A...403L...1A}, with subsequent confirmation by
H.E.S.S. \cite{2006Sci...314.1424A}, VERITAS
\cite{2008ApJ...679..397A} and MAGIC \cite{2008ApJ...685L..23A}. M~87
is a giant radio galaxy at a distance of $16.7 \pm 0.2 \U{Mpc}$,
displaying a prominent misaligned jet, with an orientation angle of
$\leq20^{\circ}$ to the line-of-sight. The mass of the central black
hole is estimated to be $\sim3\times10^9$~M$_\odot$.  The TeV source
is strongly variable, and has undergone three episodes of enhanced
emission in 2005, 2008 and 2010 (Figure~\ref{M87}). The results are
summarized by Abramoswski et al \cite{2012ApJ...746..151A}. Causality
arguments use the shortest variability timescale of around $1\U{day}$
to place strong constraints on the size of the TeV emission region,
corresponding to only a few Schwarzschild radii. The TeV emission
region cannot be directly resolved with IACTs, but correlations with
contemporaneous X-ray and radio observations provide some clues to its
location. Two structures are of particular interest: the inner region
close to the central black hole (the ``core''), and HST-1, a bright
jet feature first resolved in the optical band by the \textit{Hubble
Space Telescope}. HST-1 underwent a multi-year flare in radio, optical
and X-rays, peaking around the time when the first short-term
variability was detected at VHE energies
\cite{2006Sci...314.1424A}. In contrast to this, the 2008 VHE flare
was accompanied by enhanced radio and X-ray fluxes from the core
region \cite{2009Sci...325..444A}. The 2010 VHE flare showed no
enhanced radio emission from the core, although an enhanced X-ray flux
was observed 3 days after the VHE peak. The results, therefore, remain
somewhat ambiguous, and the possibility remains that the observed VHE
flares may have different origins, or that the total VHE emission may
be the sum of multiple components. Given the burdensome observing
requirements for instruments with a limited duty cycle, M~87
represents the best example of the importance of data-sharing and
coordinated observing planning between the various IACTs.

The closest active galaxy, Centaurus A, at a distance of $3.8\U{Mpc}$,
was identified as a VHE source in a deep, 120 hour exposure by
H.E.S.S. \cite{2009ApJ...695L..40A}. Cen~A is among the faintest VHE
sources detected, with a flux of 0.8\% of the Crab Nebula above
$250\U{GeV}$. The emission is steady, although the variability is not
strongly constrained as a result of the low flux level. As with M~87,
numerous sites for the production of the TeV emission have been
suggested, from the immediate vicinity of the central black hole (with
a mass of $\sim5\times10^7$~M$_\odot$), to the AGN jet, or even beyond
\cite{2006MNRAS.371.1705S}. The final, and most recent, addition to
the known TeV radio galaxies is NGC~1275, identified by MAGIC
\cite{2011arXiv1111.0143L}. NGC~1275 is the central galaxy of the
Perseus cluster, at a distance of $72.2\U{Mpc}$. The VHE emission has
a flux of $\sim3\%$ of the Crab Nebula above $100\U{GeV}$ and exhibits
a very soft spectrum, with a power-law index of $3.96\pm 0.37$
\cite{2011arXiv1110.5358H}, indicating a sharp turnover from the
measured Fermi spectrum at lower energies, where the index is $2.09
\pm 0.02$ \cite{2011MNRAS.413.2785B}.

\subsection{Starburst Galaxies}

Starburst galaxies are those which exhibit an extremely high rate of
star formation, sometimes triggered by interaction with another
galaxy. High cosmic-ray and gas densities in the starburst region make
these objects promising targets for gamma-ray observations, with
emission predicted to result from the interactions of
hadronic cosmic rays in the dense gas. TeV emission has now been
identified from two starburst galaxies: M~82
\cite{2009Natur.462..770V} and NGC~253 \cite{2009Sci...326.1080A}.

M~82 is a bright galaxy located at a distance of approximately
$3.9\U{Mpc}$, with an active starburst region at its centre. The star
formation rate in this region is approximately 10 times that of the
Milky Way, with an estimated supernova rate of 0.1 to 0.3 per year.  A
deep VERITAS exposure ($137\U{hours}$) in 2008-2009 resulted in a
detection of gamma-ray emission from M~82 with a flux of $(3.7 \pm
0.8_{stat} \pm 0.7_{syst}) \times 10^{-13} \UU{cm}{-2} \UU{s}{-1}$
above the 700\U{GeV} energy threshold of the analysis. NGC~253 lies at
a distance of $2.9-3.6\U{Mpc}$, and also has a central, compact
($\sim100\U{pc}$) starburst region. The supernova rate in this region
is estimated at $\sim0.03$ per year. TeV emission was detected by
H.E.S.S. with an integrated flux above $220\U{GeV}$ of $(5.5 \pm
1.0_{stat} \pm 2.8_{syst}) \times 10^{-13} \UU{cm}{-2} \UU{s}{-1}$
\cite{2009Sci...326.1080A}.

The emission from both M~82 and NGC~253 is consistent with the
predictions of models based on the acceleration and propagation of
cosmic rays in the starburst core (e.g. \cite{2011ApJ...734..107L}),
assuming that they act as efficient ``proton calorimeters''
(i.e. cosmic rays lose the majority of their energy to collisions). In
this case, the estimated cosmic ray density in the starburst region is
2-3 orders of magnitude larger than that of the Milky Way. An
alternative explanation is proposed by Mannheim et
al. \cite{2010arXiv1010.2185M}, who suggest that the TeV luminosity is
consistent with the combined emission from a large population of
pulsar wind nebulae, which result from the elevated supernova
rate. More accurate TeV spectra, and observation of other starburst
classes, such as ultra-luminous infrared galaxies, should provide more
insight in the future.

\subsection{The Large Magellenic Cloud}

Galaxies of the Local Group are also of interest to TeV observatories,
although the predicted fluxes due to cosmic ray acceleration and
propagation lie below the current instrumental sensitivity in the TeV
range. Given their proximity, the possibility arises of detectable
emission from individual objects, or from localized regions,
particularly in the Milky Way's satellite galaxies. H.E.S.S. has
recently identified the first such object in the Large Magellenic
Cloud (LMC), at a distance of $48\U{kpc}$ \cite{2011ICRC_Komin}. An
unresolved source with a flux of $\sim2\%$ of the Crab Nebula
($1.5\times10^{-12} \U{erg} \UU{cm}{-2} \UU{s}{-1}$ between 1 and 10
TeV) was detected, consistent with the location of
PSR~J0357-6910. This object is the most powerful pulsar known, with a
spin-down energy of \.{E}$=4.8\times10^{38}\U{erg}\UU{s}{-1}$. Given
the positional coincidence, and based on comparisons with similar
objects within our Galaxy, it seems likely that the TeV emission is
due to inverse Compton emission from electrons in the pulsar's wind
nebula interacting with a strong infrared target photon field. This is
in contrast to the extended GeV emission which has been observed from
the LMC by \textit{Fermi}-LAT, which is attributed to cosmic ray
acceleration and interactions in the massive star forming region of
30~Doradus \cite{2010A&A...512A...7A}.

\subsection{Other Extragalactic TeV targets}
TeV observations have also been used to place important constraints on
the gamma-ray emission from numerous undetected extragalactic source
classes, including galaxy clusters and potential sources of ultra-high
energy cosmic rays. Here we summarize two of the most important:
gamma-ray bursts, and the predicted sites of dense regions of dark
matter particles.

\noindent {\bf Gamma-Ray Bursts:} GRBs are the signatures of brief, extremely
energetic explosions which occur at cosmological distances. They are
observationally divided into short and long classes, which are
presumably the result of different progenitor systems. Long duration
GRBs are generally ascribed to the collapse of massive, rapidly
rotating stars into black holes. The origin of the short bursts is
less certain, although neutron star - neutron star merger events are
among the favoured candidates. The search for $>100\U{GeV}$ emission
has been a long-running goal of both the IACTs and particle detectors
(see \cite{2009astro2010S.316W} for a review). Given the brief
duration of GRB emission, the wide field-of-view of particle detectors
is particularly important in this regard, although observations so far
have been hampered by limited sensitivity. IACTs, conversely, must be
re-pointed rapidly on receipt of an alert. The detection of delayed
high energy emission by \textit{Fermi}-LAT , lasting hundreds to
thousands of seconds longer than the sub-MeV emission, has provided
additional impetus to the search \cite{2009Sci...323.1688A,
2009ApJ...707..580A}, and IACTs now regularly target burst locations
within $<100\U{s}$ of the burst alert. The task is difficult, since
the burst must also be at a small enough redshift such that the high
energy emission is not completely suppressed by photon-photon pair
production ($z\lesssim0.5$). No convincing signals have been detected
as yet (e.g. \cite{2010ApJ...720.1174A, 2010A&A...517A...5A, 2009A&A...495..505A,
2011arXiv1109.0050V}), although predictions based on the brightest
bursts observed by the LAT indicate that the potential for detecting
TeV emission associated with a GRB is promising, assuming no intrinsic
spectral cut-off of the high energy emission
\cite{2011arXiv1111.1326A}.

\begin{figure}[!t]
  \vspace{0mm} \centering \includegraphics[width=\columnwidth,clip,trim = 1mm
  1mm 0mm 1mm]{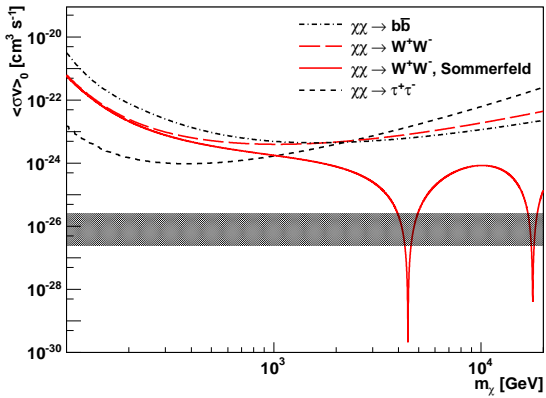}
  \caption{Upper limits at the 95\% confidence level on the
  velocity-weighted annihilation neutralino cross-section for
  different annihilation channels, based on VERITAS observations of
  the Segue I dwarf spheroidal galaxy. The dark band represents the
  typical range of predictions. Figure from \cite{2011arXiv1110.6615V}}
  \label{DM}
 \end{figure}

\noindent {\bf Extragalactic Dark Matter:} The search for the
self-annihilation signature of dark matter particles in astrophysical
objects is wide-ranging, and complementary to direct detection
techniques on the Earth (see \cite{2011ARA&A..49..155P} for an
excellent review). As discussed later in this review, the centre of
our own Galaxy is a natural target for dark matter searches, and
provides the most stringent limits to date
\cite{2011PhRvL.106p1301A}. Objects outside of our own Galaxy are also
worthy of investigation, however, and are potentially much less
affected by contamination from unknown astrophysical background
sources (supernova remnants, pulsar wind nebulae, etc.). Dwarf
spheroidal galaxies of the local group are among the most promising of
these, due to their proximity and their presumed large dark matter
content. The Sloane Digital Sky Survey has more than doubled the known
population of dwarf spheroidals, providing additional targets for the
TeV searches. No signals have been detected as yet, despite deep
exposures on a number of objects \cite{2011arXiv1110.6615V,
2011JCAP...06..035A, 2011APh....34..608H, 2009ApJ...691..175A,
2008APh....29...55A}. Figure ~\ref{DM} shows upper limits on the
annihilation cross-section derived for various annihilation channels
using VERITAS observations of the Segue~I dwarf spheroidal.

\section{Galactic TeV Sources}

\begin{figure}[!t]
  \vspace{0mm} \centering \includegraphics[width=\columnwidth]{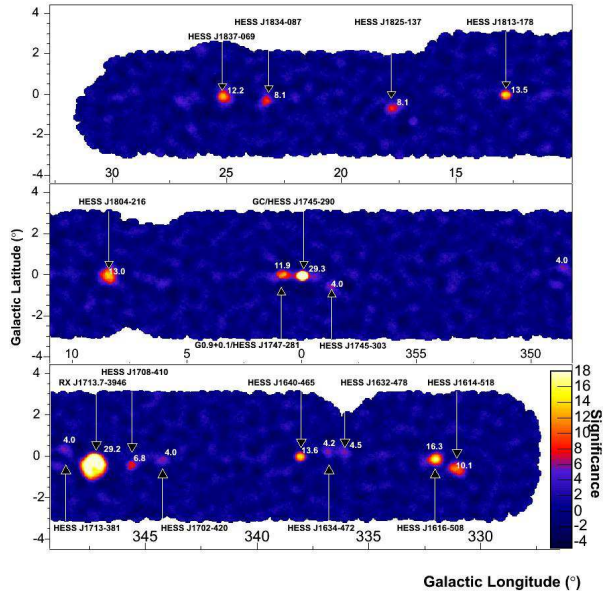}
  \caption{Significance map of the Galactic Plane from the original
  H.E.S.S. survey in 2004, based on $230\U{hours}$ of
  observations. The complete survey now extends over the range from
  $l=280^{\circ}$ to $60^{\circ}$ and $b=-3.5^{\circ}$ to
  $+3.5^{\circ}$, and comprises $2300\U{hours}$ of observations \cite{2011ICRC_Gast}.
  Figure from \cite{2006ApJ...636..777A}}
  \label{plane}
 \end{figure}

There are presently $\sim80$ known TeV sources within our Galaxy, as
indicated either by their association with known Galactic sources at
other wavelengths, or by their location in the Galactic plane - a
particularly compelling argument when coupled with a resolvable
angular extent. The majority of these sources were identified as TeV
emitters during the H.E.S.S. survey of the inner Galaxy
\cite{2006ApJ...636..777A}. H.E.S.S. is the only IACT currently
operating in the southern hemisphere, which allows it to view the
inner Galaxy at high elevation, and hence with a low energy threshold
and good sensitivity. H.E.S.S. was the first instrument with
sufficient sensitivity to observe sources with $\lesssim10\%$ of the
Crab Nebula flux in this region, and the results have been revelatory
- the Galactic plane is densely populated with TeV sources, primarily
clustered within the inner $\pm60^{\circ}$ in Galactic longitude
(Figure~\ref{plane}). The most recent survey results consist of 2300
hours of observations, allowing the detection of over 50 sources
within the range $l=280^{\circ}$ to $60^{\circ}$ and $b=-3.5^{\circ}$
to $+3.5^{\circ}$ \cite{2011ICRC_Gast}. Observations of the outer
Galactic regions by VERITAS, MAGIC, Milagro and ARGO-YBJ have revealed
a less densely populated sky, but containing some unique objects of
particular interest for TeV studies. Many Galactic TeV sources are
extended, allowing detailed studies of source morphology and spatially
resolved spectra, while others are time variable and/or periodic. The
various source classes are discussed in some detail below.

\subsection{The Galactic Centre and Ridge}
\begin{figure}[!t]
  \vspace{0mm} \centering \includegraphics[width=\columnwidth]{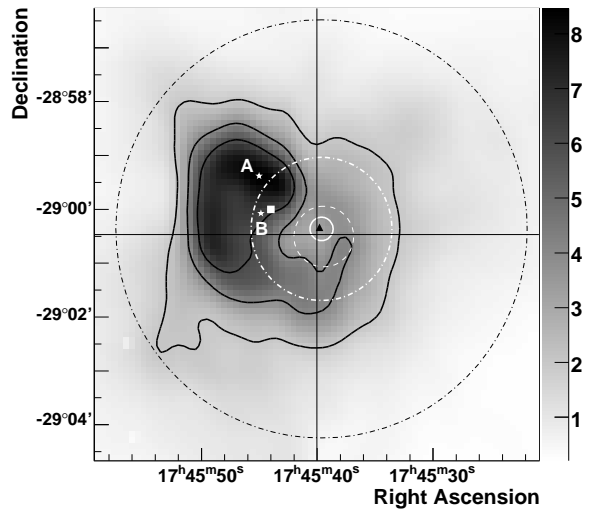}
  \caption{90~cm VLA radio flux density map of the innermost
    $20\U{pc}$ of the Galactic Centre, showing emission from the SNR
    \aeast.  Black contours denote radio flux levels; the centre of
    the SNR is marked by the white square, and the positions of
    \astar\ and \pwn\ are given by the cross hairs and the black
    triangle, respectively. The 68\% CL total error contour of the
    best-fit centroid position of \hgc\ is given by the white
    circle. Figure from \cite{2010MNRAS.402.1877A}: see that paper for
    full details.}
  \label{GC}
 \end{figure}

A TeV source at the location of the Galactic Centre has been reported
by various IACTs \cite{2004ApJ...606L.115T, 2004ApJ...608L..97K,
2006ApJ...638L.101A, 2006PhRvL..97v1102A}. Determining the nature of
this source is a difficult task, due to the complexity of the region,
which includes multiple different potential counterparts. The most
detailed studies have been performed by H.E.S.S., which reveal that
the emission is dominated by a bright central source, HESS~J1745-290,
lying close to the central supermassive black hole, \astar.  An
additional, fainter, component is also seen, which extends in both
directions along the Galactic plane \cite{2006Natur.439..695A}. The
extended component is spatially correlated with a complex of giant
molecular clouds in the central $200\U{pc}$ of the Milky Way, and the
TeV emission can be attributed to the decay of neutral pions produced
in the interactions of hadronic cosmic rays with material in the
clouds. The central source is point-like, steady and exhibits a curved
power-law spectrum \cite{2009A&A...503..817A}. Its location with
respect to three of the most likely counterparts is shown in
Figure~\ref{GC}. This study reveals that the source centroid is
displaced from the radio centroid of the supernova remnant Sgr A East,
excluding this object with high probability as the dominant source of
the VHE gamma-ray emission, and leaving \astar and the pulsar wind
nebula \pwn~as the most likely counterparts
\cite{2010MNRAS.402.1877A}.

The Galactic Centre is also a prime candidate region in which to
search for gamma-ray emission due to dark matter particle
self-annihilation. The analysis is complicated, however, because of
the high background due to astrophysical sources. An analysis by
H.E.S.S. using an optimized background subtraction technique shows no
hint of a residual dark matter gamma-ray flux at a projected distance
of $r\sim45-150\U{pc}$ from the Galactic Centre
\cite{2011PhRvL.106p1301A}.

\subsection{The Crab Nebula and Pulsar}

The Crab is the nearby ($2.0\pm0.2\U{kpc}$) remnant of a historical
supernova explosion, observed in 1054 A.D. There is no detected shell,
and the broadband emission below $\sim100\U{MeV}$ is dominated by a
bright synchrotron nebula, powered by a central pulsar
(PSR~B0531+21). PSR~B0531+21 is the most energetic pulsar in
our Galaxy, with a pulse period of $33\U{ms}$, and a spin-down power
of $4.6\times10^{38}\U{erg}\UU{s}{-1}$. The Crab Nebula and Pulsar
hold a unique place in the development of TeV astronomy: the birth of
the field as an astronomical discipline can be traced to the detection
of the Crab Nebula TeV source by Weekes et al. using the Whipple
$10\U{m}$ telescope, in the first application of the imaging
atmospheric Cherenkov technique
\cite{1989ApJ...342..379W}. Subsequently, the Crab has acted as a
bright, standard candle for TeV observatories.

\begin{figure*}[!t]
  \centering 
  \includegraphics[height=1.9in]{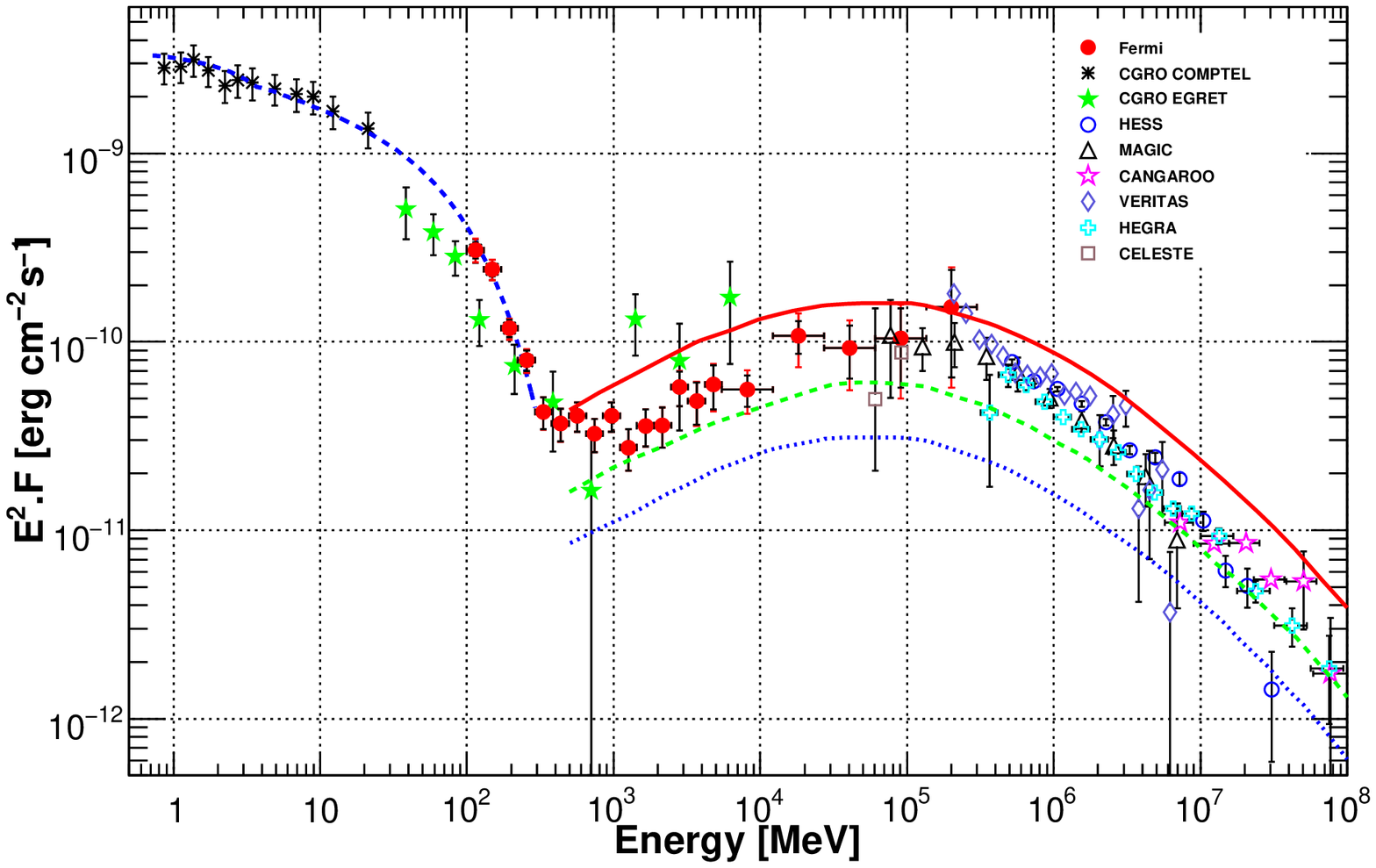}\includegraphics[height=1.85in]{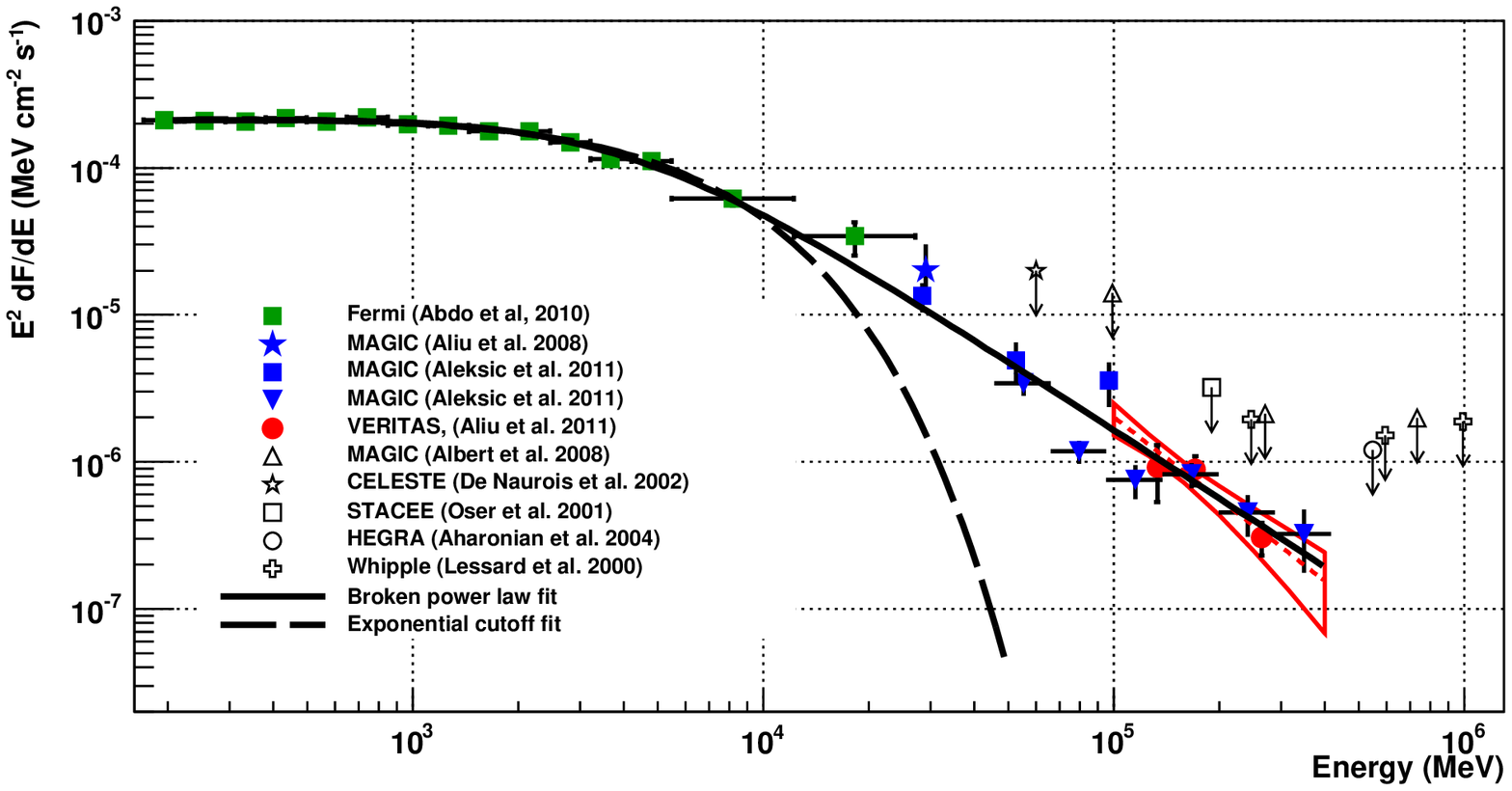}
  \caption{ {\bf Left:} The spectral energy distribution of the Crab
    Nebula from Abdo et al. \cite{2010ApJ...708.1254A}. Their fit to
    the synchrotron component is shown (blue dashed line), as well as
    inverse Compton spectra from Atoyan and Aharonian
    \cite{1996MNRAS.278..525A} for assumed magnetic field strengths of
    $100\U{\mu G}$ (solid red line), $200\U{\mu G}$ (dashed green
    line) and $300\U{\mu G}$ (dotted blue line). {\bf Right:} High
    Energy spectrum of the Crab Pulsar. The black dashed line shows a
    fit of a power law with exponential cut-off to the
    \textit{Fermi}-LAT data alone; the solid line shows a broken power
    law fit to the LAT and IACT data. The red dashed line and bowtie
    shows a power law fit to the VERITAS points alone. Figure courtesy
    of N. Otte (priv. comm.).}
    \label{crab}
\end{figure*}

The SED of the non-thermal nebula emission displays two components
(Figure~\ref{crab}). The dominant, low frequency component is
explained by synchrotron radiation of high energy electrons spiraling
in the magnetic field of the nebula \cite{1984ApJ...283..710K,
1996ApJ...457..253D}. The higher frequency component is attributed to
inverse Compton scattering of lower energy photons by these electrons,
including microwave background photons, far infrared and the
electron-synchrotron photons themselves. The electron population
reaches energies of at least $10^{15}\U{eV}$, through acceleration
occuring both in a relativistic particle outflow driven by the
spin-down energy of the pulsar, and in shocks where this outflow
encounters the surrounding nebula. The highest energy particles likely
require an alternative explanation for their origin, such as direct
acceleration in intense electric fields associated with the pulsar
itself \cite{2011Sci...331..739A}. Observations of the synchrotron
nebula from radio to X-ray wavelengths provide high resolution imaging
of the emission region; however, the synchrotron data alone only
contain information concerning the product of the magnetic field
strength and the relativistic electron density. Since the inverse
Compton component is independent of the magnetic field strength, the
combined SED allows an estimate of the Nebula magnetic field, which is
now constrained to be between 100 and $200\U{\mu G}$
\cite{2010ApJ...708.1254A}.

The search for a VHE component to the pulsed emission from the Crab
has been long and, until recently, fruitless. Despite discouraging
model predictions, and the detection of spectral cut-offs below
$10\U{GeV}$ in other pulsars, the fact that no super-exponential
cut-off was observed by EGRET in the Crab Pulsar spectrum initially
provided some encouragement for a continued search by IACTs
\cite{2001A&A...378..918K}. \textit{Fermi}-LAT subsequently extended
the GeV spectrum and measured a sharp spectral cut-off at $6\U{GeV}$
\cite{2010ApJ...708.1254A}. A campaign by MAGIC, using a specially
designed ``analog sum'' hardware trigger, provided the first
ground-based measurement of gamma-ray emission from the Crab pulsar
\cite{2008Sci...322.1221A}. The initial MAGIC flux measurement above
$25\U{GeV}$ was, like the \textit{Fermi}-LAT result, consistent with
an exponential cut-off. The existence of an exponential cut-off is a
natural consequence of emission due to curvature radiation, as favored
by various models (e.g. \cite{2008ApJ...676..562T}). Both VERITAS and
MAGIC recently presented new results, which challenge this paradigm
\cite{2011Sci...334...69V, 2011ApJ...742...43A,
2011arXiv1109.6100S}. Pulsed emission is observed to extend up to well
beyond $100\U{GeV}$, and the combined LAT-IACT spectrum can best be
fit with a broken power law. The explanation for this high energy
component is an open question, at present. Gamma-ray opacity arguments
require that the emission zone of the highest energy photons must be
at least 10 stellar radii from the surface of the neutron star - much
further than previously assumed. The results require either a
substantial revision of existing models of high energy pulsar
emission, or the addition of a new component, not directly related to
the MeV-GeV emission.

As mentioned above, the Crab has been used as a standard candle in TeV
astronomy, on the assumption that its emission was steady. This is now
demonstrably false, at least at energies below $\sim1\U{GeV}$, with
the detection of multiple day-scale flaring events
\cite{2011ApJ...741L...5S, 2011Sci...331..739A}, and long term
variation in the hard X-ray/ soft gamma-ray regime
\cite{2011ApJ...727L..40W}. The GeV spectrum during flares indicates
that the emission is confined to the synchrotron component of the SED,
a conclusion which is supported by the rapid timescale of the events
(since the inverse Compton or Bremsstrahlung cooling time of the
emitting electrons is much greater than the observed flare
duration). At higher energies, some evidence for an enhanced flux
during HE flare states has been presented by ARGO-YBJ
\cite{2011arXiv1107.3404D}. IACT measurements do not support these
results, but are not necessarily in conflict, given the differing duty
cycles. Detailed measurements with IACTs during future flare states
are required to resolve this question.

\subsection{Pulsar Wind Nebulae}

Pulsar wind nebulae are the most abundant class of known VHE emitters
in the Galaxy, with $\sim30$ firm examples, and numerous other sources
where the PWN association is more tentative (for reviews see
e.g. \cite{2011AIPC.1357..213D, 2010AIPC.1248...25K,
2011heep.conf..373S}). The essential emission mechanisms - shock
accelerated leptons producing synchrotron and inverse Compton
radiation - have already been described for the case of the Crab PWN,
but the Crab is far from the typical object. Understanding of the
structure and evolution of PWN has advanced significantly over the
past few years, in particular thanks to the high resolution X-ray
imaging provided by \textit{Chandra} (see Gaensler and Slane
\cite{2006ARA&A..44...17G} for a detailed review). Initially, the PWN
expands uniformly from the central pulsar, while at later stages the
nebula may be confined and distorted by the reverse shock from the
expanding supernova remnant (SNR).

\begin{figure}[!t]
  \vspace{0mm} \centering \includegraphics[width=\columnwidth]{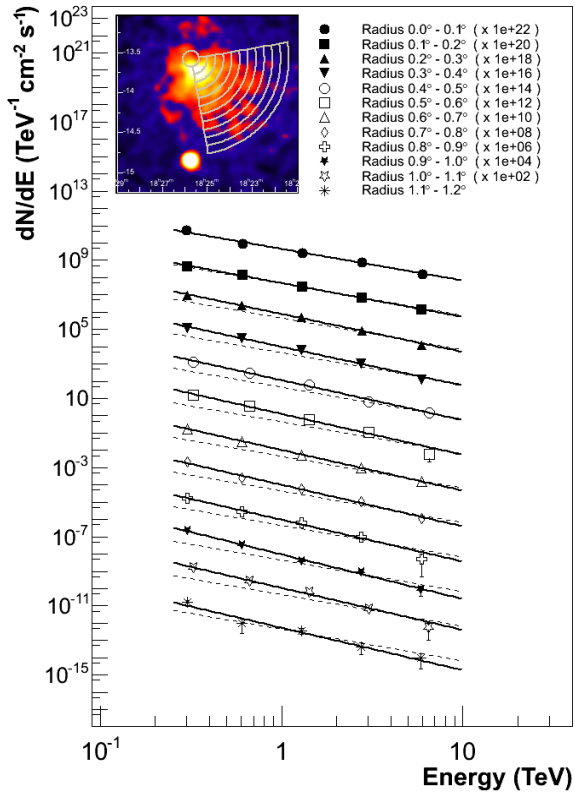}
  \caption{ Inset: H.E.S.S. gamma-ray excess map for HESS J1825-137. The
    wedges show the radial regions with radii in steps of
    $0.1^{\circ}$ in which the energy spectra were determined. The
    main figure shows the differential energy spectra for the regions
    illustrated in the inset, scaled by powers of 10 for the purpose
    of viewing. The spectrum for the analysis at the pulsar position
    is shown as a reference along with the other spectra as dashed
    line. Figure from \cite{2006A&A...460..365A}: see that paper for
    full details.}
  \label{PWN}
 \end{figure}

At TeV energies, young PWN are usually still embedded within their
parent SNR and are point-like, within the angular resolution of
IACTs. They are positionally coincident with a bright X-ray
synchrotron nebula powered by a pulsar with very high spin-down
luminosity (e.g. G0.9+0.1 \cite{2005A&A...432L..25A}, HESS~J1813-178
\cite{2005Sci...307.1938A}, G54.1+0.3
\cite{2010ApJ...719L..69A}). More evolved PWN, with ages $>10,000$
years, are usually much larger, and their TeV emission can be
spatially resolved and mapped. The pulsar powering the nebula is often
offset from the center of the TeV emission, probably for reasons
related to density gradients in the medium surrounding the SNR
\cite{2001ApJ...563..806B}. Remarkably, the TeV nebulae are often two
or three orders of magnitude larger than the corresponding X-ray PWN,
and the TeV PWN sizes tend to increase with age, while the X-ray PWN
sizes show the opposite trend. This can be understood as a result of
the fact that the electron population which is responsible for the TeV
inverse Compton flux has lower energies than the electrons which
produce the X-ray synchrotron emission. They therefore cool more
slowly, and survive for longer so, while the X-ray nebula is dominated
by freshly accelerated particles, the TeV nebula can record the entire
history of particle propagation away from the pulsar. A natural result
of this is that the observed TeV spectrum should vary with distance
from the pulsar. One of the best examples of this is HESS J1825-137,
associated with the PWN of the pulsar PSRJ1826-1334
\cite{2006A&A...460..365A}. Figure~\ref{PWN} shows the spatially
dependent spectra for this source, which soften with increasing
distance from the pulsar. This is interpreted as the natural effect of
both inverse Compton and synchrotron cooling of the electron
population during propagation. A counter-example is the case of Vela-X
\cite{2006A&A...448L..43A, 2011ApJ...743L...7H}, in which no spectral
variability is seen over the extended nebula, suggesting that cooling
does not play an important role.

\subsection{Supernova Remnants}

The search for the origin of the cosmic rays triggered the development
of gamma-ray astronomy, and continues to motivate many gamma-ray
observations. Chief among these is the study of supernova remnants,
which are believed to efficiently accelerate particles at the shock
front where the expanding SNR encounters the surrounding medium
(e.g. \cite{1994A&A...287..959D}). This likely occurs through
diffusive shock acceleration (first order Fermi acceleration), in
which charged particles are reflected from magnetic inhomogeneities
and repeatedly cross the shock front, gaining energy with each
crossing (see e.g. \cite{2001RPPh...64..429M,
1987PhR...154....1B}). As well as plausibly providing enough energy to
explain the observed Galactic cosmic ray population, this process
naturally produces a power law distribution of particle energies with
an index of $\sim2$, which matches the cosmic ray spectrum (after
accounting for diffusion and escape). In recent years, the importance
of magnetic field amplification by the accelerated particles
themselves has been increasingly recognised, and plays a particular
role in explaining the existence of the highest energy Galactic cosmic
rays, around the cosmic ray knee region (at
$\sim3\times10^{15}\U{eV}$).

The evidence for efficient leptonic acceleration in SNRs is now
clearly established (e.g. \cite{2007Natur.449..576U}); however, the
question of whether SNR are efficient hadron accelerators is more
difficult to answer. A definitive measurement would be the detection
of high energy neutrinos from an SNR, but the expected fluxes are
likely below the sensitivity thresholds of current neutrino
observatories. Gamma-ray observations may provide the key, since the
interactions of high energy nuclei with target material produce
neutral pions, which decay immediately into gamma-rays. Disentangling
the spectral signature of this process from other sources of gamma-ray emission
(i.e. leptonic inverse Compton and bremsstrahlung processes) is
difficult, but not impossible. Two classes of gamma-ray source are of
interest for these studies: those which can be clearly associated with
SNR shells, based on the gamma-ray morphology, and sources which are
coincident with a massive volume of target material, such as molecular
clouds.

\begin{figure*}[!t]
  \centering 
  \includegraphics[height=1.95in]{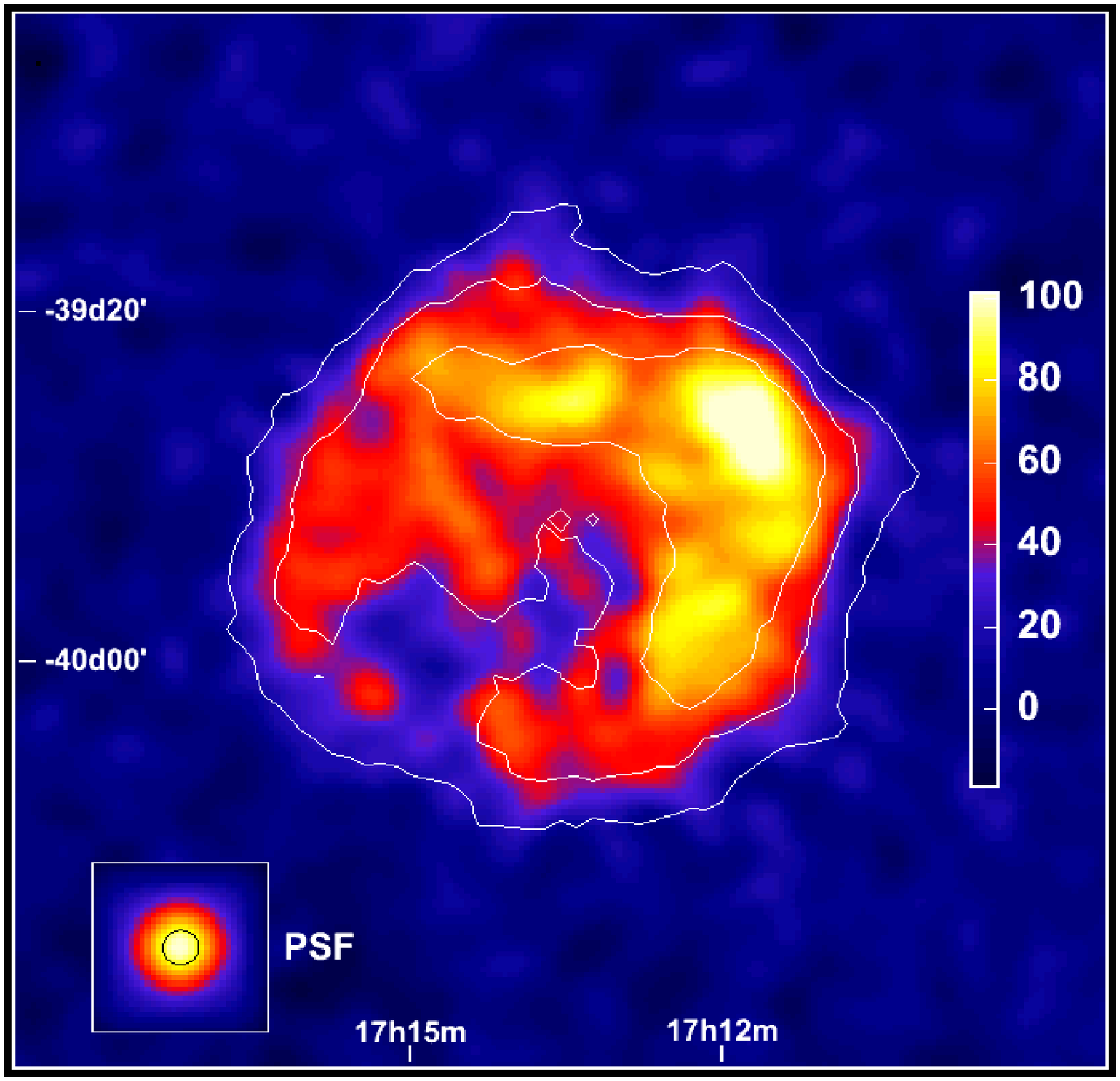}\includegraphics[height=2.2in]{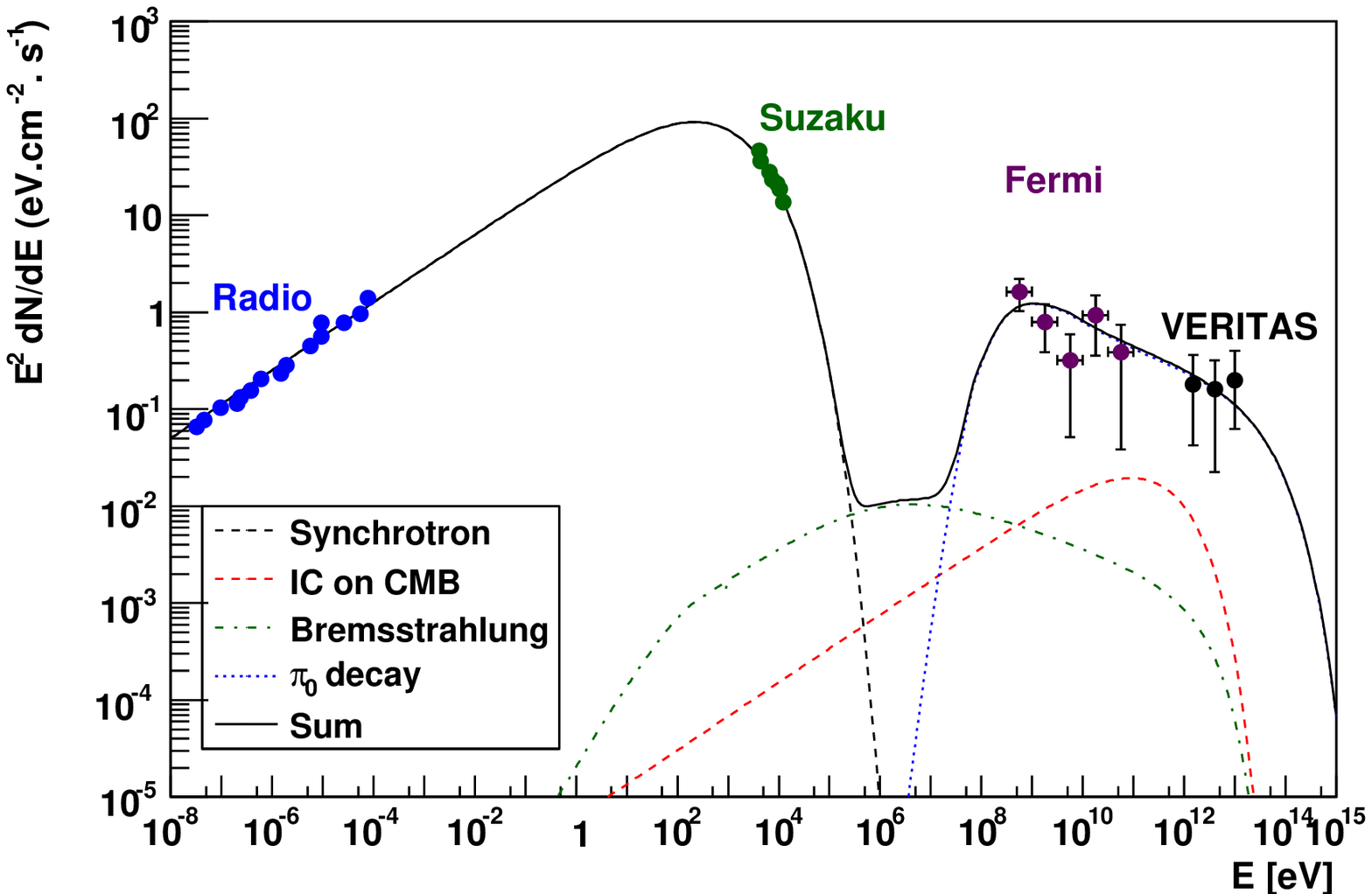}
  \caption{ {\bf Left:} H.E.S.S. map of gamma-ray excess events for RXJ 1713.7-3946 -
    the first SNR shell to be resolved at TeV energies. Figure from
    \cite{2006A&A...449..223A}. {\bf Right:} The broadband SED of
    Tycho's SNR from \cite{2012ApJ...744L...2G}, together with models
    for the various emission components (dominated by hadronic
    processes in the gamma-ray band). See paper for details.}
    \label{SNR}
\end{figure*}

The definitive association of gamma-ray emission with an SNR shell is
often difficult to make, due to the presence of other potential
counterparts, particularly PWN. A handful of shell-type SNRs have been
unequivocally identified as gamma-ray sources by IACTs. This
identification can be made on the basis of the observed shell
morphology, (RXJ~1713.7-3946 \cite{2006A&A...449..223A},
RXJ~0852.0-4622 (Vela Jr) \cite{2007ApJ...661..236A}, HESS~J1731-3467
\cite{2011A&A...531A..81H}, SN1006 \cite{2010A&A...516A..62A} and,
possibly, RCW~86 \cite{2009ApJ...692.1500A}), or, in the case of
Tycho's SNR \cite{2011ApJ...730L..20A}, on the positional coincidence,
coupled with the fact that the progenitor was a known Type Ia
exposion, and so no compact object is present. Figure~\ref{SNR} shows
the gamma-ray map for the first SNR shell to be resolved,
RXJ~1713.7-3946. The recent addition of \textit{Fermi}-LAT
observations to the broadband spectrum of RXJ~1713.7-3946
\cite{2011ApJ...734...28A} are consistent with a leptonic origin as
the dominant mechanism for the gamma-ray emission. A counter-example
is illustrated by the spectrum in Figure~\ref{SNR}, which corresponds
to Tycho's SNR. In this case, the fact that the broadband gamma-ray
spectrum can be fit with a single hard power law from $500\U{MeV}$ to
$10\U{TeV}$ favours a hadronic origin \cite{2011arXiv1105.6342M,
2012ApJ...744L...2G}. This interpretation is not completely
compelling, however, given the large statistical errors in the
measurements, and the impact of various unknown parameters such as the
SNR distance, and possible enhancements of the gamma-ray flux due to a
nearby molecular cloud. Additionally, Atoyan and Dermer
\cite{2011arXiv1111.4175A} describe a two-zone leptonic model which
provides an acceptable spectral fit. Future measurements of the
spectrum below $500\U{MeV}$, and deeper exposure at TeV energies, will
further test the differing interpretations.

The intensity of gamma-ray emission due to hadronic interactions
depends upon the flux of high energy nuclei, and also upon the density
of target material. Regions of high matter density (e.g. molecular
clouds with densities $>100\UU{cm}{-3}$), situated close to sites
of particle acceleration (such as SNRs), can therefore be expected to
produce a large gamma-ray flux due to hadronic interactions.  At TeV
energies, $\sim10$ likely candidates for this process have been
identified. The task of identification is complicated, both for the
usual reasons of source confusion, and also because the evidence for a
molecular cloud / SNR interaction is only definitive in those cases
where the cloud morphology is visibly deformed by the expanding SNR,
and/or where sites of hydroxyl (OH) maser emission indicate the
presence of shocked molecular material. One of the best examples of
this source class is the old remnant W28
\cite{2008A&A...481..401A}. H.E.S.S. observations of this region show
four distinct sites of emission, with three of the four showing a
resolvable angular extent ($\sim10'$). Each of the TeV sources is
positionally coincident with a molecular cloud. Assuming that the TeV
emission is due to hadronic cosmic rays interacting with the cloud
material, the cosmic ray density is inferred to be a factor of 10 to
30 times greater than in the solar neighbourhood.

\subsection{Star forming regions}

The process of diffusive shock acceleration is not limited to
supernova remnant shells. An alternative scenario invokes particle
acceleration at the shock formed by the collision between the
supersonic stellar winds of massive stars in close binary
systems. Stellar winds may also become collectively important in large
assemblies of massive stars.  The combined effect of the stellar
winds, coupled with the effect of multiple SNRs, results in an overall
wind from the cluster which forms a giant \textit{superbubble} in the
interstellar medium. Particle acceleration can occur where the cluster
wind interacts with the surrounding medium
(e.g. \cite{2009Natur.460..701B}).

Massive star associations are naturally likely to host other potential
source counterparts for TeV emission, such as compact object binary
systems, individual supernova remnants and pulsar wind nebulae. A case
in point is the young, open cluster Westerlund 2, containing the
Wolf-Rayet binary system WR~20a.  This was originally suggested as a
plausible counterpart to the unidentified source HESS~J1023-575,
with the emission presumed to be connected to either the Wolf-Rayet
binary, or the combined cluster wind \cite{2007A&A...467.1075A}. A
re-assesment of this region, informed by a deeper H.E.S.S. exposure
and results from \textit{Fermi}-LAT, alters the picture
\cite{2011A&A...525A..46H}. The LAT detects an energetic pulsar,
PSR~J1022-5746, which drives a PWN which is bright in GeV gamma-rays
\cite{2011ApJ...726...35A}. Given the ubiquity of bright TeV PWN, this
now seems the most likely explanation for the TeV source. A similar
conclusion may arise for the first TeV source to be linked with a
massive star association, TeV~2032+4130 (coincident with the Cyg OB2
association). In this case, the LAT pulsar (PSR J2032+4127) is
sufficiently energetic to explain the TeV emission, although no PWN has
been detected as yet.

Other unidentified sources which have been linked with massive star
clusters and associations include HESS~J1646-458 (Westerlund 1)
\cite{2011arXiv1111.2043H}, HESS J1614-518 (Pismis 22)
\cite{2010ASPC..422..265O}, HESS J1848-018 (W43, which hosts
Wolf-Rayet star WR121a) \cite{2008AIPC.1085..372C} and W49A
\cite{2011arXiv1104.5003B}, a massive star forming region. For all of
these, however, the evidence that particle acceleration in stellar
winds is the driving force behind the gamma-ray emission is not
definitive (e.g. \cite{2011arXiv1109.4733L}).

\subsection{Compact Object Binary Systems}

Despite many early unconfirmed claims, the first definitive detection
of a TeV gamma-ray binary system was not published until 2005. The
population has grown slowly, and now consists of four clearly
identified systems, plus marginal evidence for transient emission
associated with Cyg X-1 \cite{2007ApJ...665L..51A}. The gamma-ray
emission from binaries is believed to be powered either by accretion
(most likely onto a black hole), or by a pulsar wind. In the case of
accretion, particle acceleration takes place in relativistic jets
(e.g. \cite{2009IJMPD..18..347B}). In the pulsar wind scenario, the
acceleration occurs either in shocks where the pulsar wind encounters
the circumstellar environment (e.g. \cite{2006A&A...456..801D}), or
possibly within the pulsar wind zone itself
\cite{2008APh....30..239S}. The detection of Cyg X-1, if confirmed,
would be extremely important, since there is no doubt that this system
hosts a black hole. This is in contrast to all of the other TeV
binaries, in which the compact object is either known to be, or may
be, a neutron star.  Here we briefly summarize the results for each of
the four well-studied objects.


\noindent {\bf PSR~B1259-63/LS~2883:} This was the first gamma-ray
binary system to be firmly detected at TeV energies, and the first
known variable VHE source in our Galaxy
\cite{2005A&A...442....1A}. The system comprises a $48\U{ms}$ pulsar
orbiting a massive B2Ve companion. The orbit is highly eccentric
($e=0.87$), with a period of 3.4 years. The TeV emission exhibits two
peaks, approximately 15 days before and after periastron. Various
authors have attempted to explain the double bumped VHE lightcurve
within a 'hadronic disk scenario', in which a circumstellar disk
provides target material for accelerated hadrons, leading to $\pi^0$
production and subsequent TeV gamma-ray emission
\cite{2004ApJ...607..949K, 2006MNRAS.367.1201C}. The 2007
H.E.S.S. observations disfavour this, since the onset of TeV emission
occurs $\sim50\U{days}$ prior to periastron, well before interactions
with the disk could be expected to play a significant role
\cite{2009A&A...507..389A}. Leptonic scenarios have also been
discussed in e.g. Kangulyan et al. \cite{2007MNRAS.380..320K}.

The recent discovery of an extended and variable radio structure in
PSR~B1259-63/LS~2883 at phases far from periastron provides definitive
evidence that non-accreting pulsars orbiting massive stars can produce
variable and extended radio emission at AU scales
\cite{2011ApJ...732L..10M}. This is important, since similar
structures in LS~5039 and LS~I~+61$^{\circ}$303, where the nature of
the compact object is not certain, have been used to argue for the
existence of jets driven by accretion onto a black hole.

\begin{figure}
\centering
\includegraphics[width=\columnwidth, trim=0 0 0 0.84\columnwidth, clip=true]{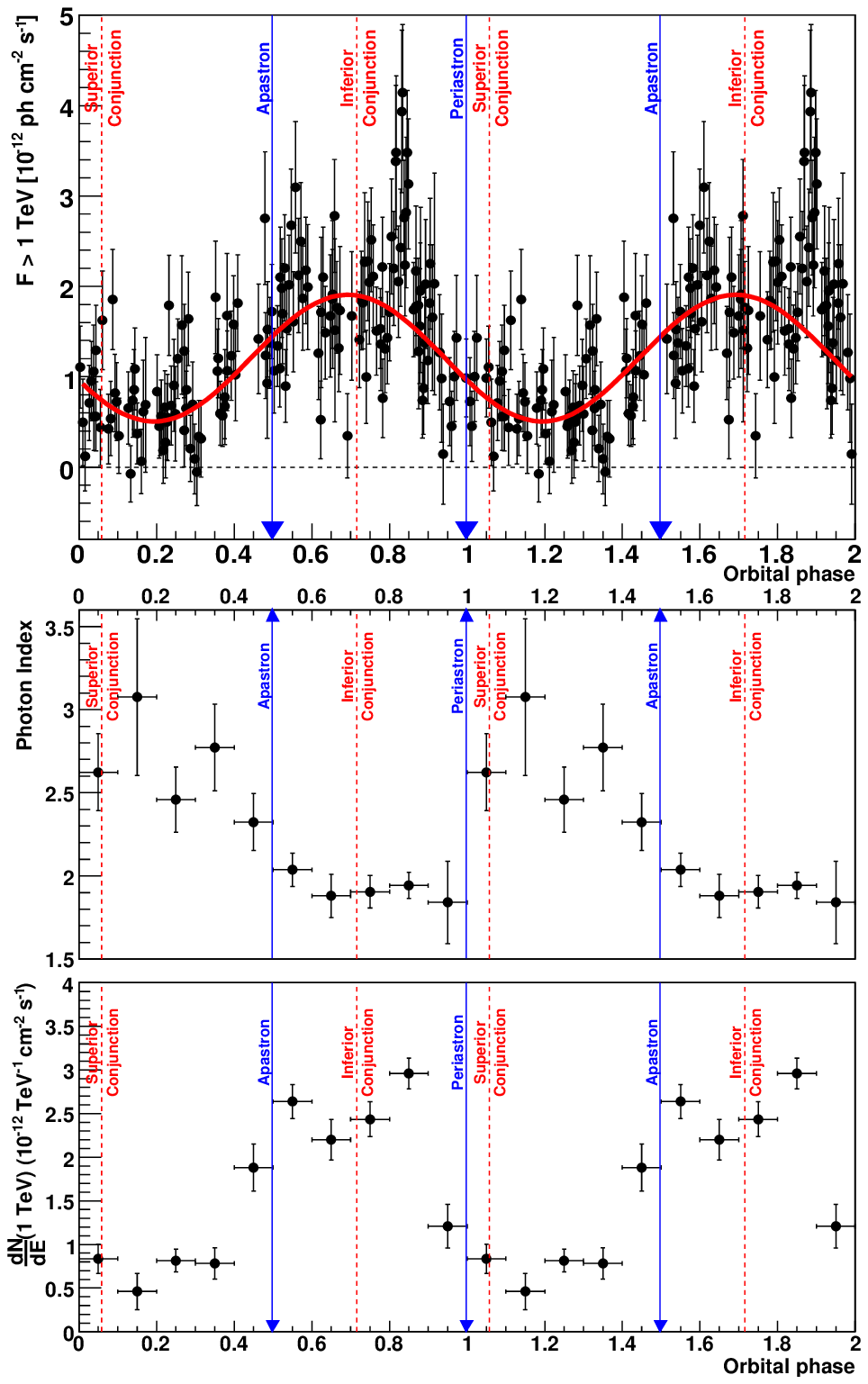}
\caption{The H.E.S.S. flux (bottom) and photon index (top) for LS~5039 as a function
of orbital phase. Figure from Aharonian et al. \cite{2006A&A...460..743A}.}
\label{HESS_LS5039_lc}
\end{figure}

\noindent {\bf LS~5039:} LS~5039 consists of a compact object, either
neutron star or black hole, orbiting a massive O6.5V
($\sim$23~M$_\odot$) star in a 3.9 day orbit. Observations by
H.E.S.S. in 2004 revealed that LS~5039 is a bright source of VHE
gamma-rays \cite{2006A&A...460..743A}. Unlike PSR~B1259-63 (and, to a
lesser extent, LS~I~+61${^\circ}$303) LS~5039 is almost perfectly
suited to TeV observations, with a short orbital period and a
convenient declination angle, allowing sensitive observations at all
phases over numerous orbits. The VHE emission measured by H.E.S.S. is
modulated at the orbital period, peaking around inferior conjunction,
when the compact object is closest to us and co-aligned with our
line-of-sight (Figure~\ref{HESS_LS5039_lc}). The spectrum is also
orbitally modulated, appearing significantly harder around inferior
conjuction
($\Gamma=1.85\pm0.06_{\mathrm{stat}}\pm0.1_{\mathrm{syst}}$), but with
an exponential cut-off at $E_o=8.7\pm2.0\U{TeV}$.  At GeV energies,
the source is detected by \textit{Fermi}-LAT at all orbital phases,
with the emission peaking close to \textit{superior} conjuction, in
apparent anti-phase with the VHE results \cite{2009ApJ...706L..56A}. A
sharp spectral cut-off at $E_o=1.9\pm0.5\U{GeV}$ is observed in the
LAT data near superior conjunction, indicating that the VHE spectra
cannot be simply a smooth extrapolation of the lower energy emission.


\noindent {\bf LS~I~+61${^\circ}$303:} Similar to LS~5039,
LS~I~+61${^\circ}$303 consists of a compact object, either neutron
star or black hole, in this case orbiting a B0Ve star with a
circumstellar disk ($\sim$12.5~M$_\odot$) in a 26.5 day orbit. The
detection of a variable VHE source at the location of
LS~I~+61${^\circ}$303 with MAGIC \cite{2009ApJ...693..303A}, later
confirmed by VERITAS \cite{2008ApJ...679.1427A}, established this
source as a gamma-ray binary. The object is now one of the most
heavily observed locations in the VHE sky, with deep exposures by the
two observatories spread over half a decade. Despite this, the VHE
source is much less well-characterized than LS~5039, owing to its
relatively weak VHE flux, and an inconvenient orbital period which
closely matches the lunar cycle, making observations over all orbital
phases almost impossible within a single observing season. VHE
emission was originally detected close to apastron, between phases
$\phi=0.5-0.8$. In contrast to this, VHE observations between 2008 and
2010 showed that, at least during the orbits when the source was
observed, the apastron flux was much lower than during the previous
detections \cite{2011ApJ...738....3A, 2011arXiv1111.6572M}. The
detection of TeV emission by VERITAS during a single episode close to
superior conjunction complicates the picture even further. As with
LS~5039, the \textit{Fermi}-LAT GeV emission peaks closer to
periastron, and the spectrum displays sharp cut-off, at
$E_o=6.3\pm1.1\U{GeV}$. Long term variability in the GeV band has also
been observed \cite{2012arXiv1202.1866H}.

{\bf HESS~J0632+057:} This TeV source was serendipitously detected
during H.E.S.S. observations of the Monoceros Loop SNR region, and
noted as a potential binary primarily because of its small angular
extent \cite{2007A&A...469L...1A}. Subsequent observations revealed it
to be a variable TeV source \cite{2009ApJ...698L..94A}, and co-located
with a variable radio and X-ray source, at the position of a massive
Be star, MWC\,148 \cite{2009ApJ...690L.101H}. X-ray observations with
Swift recently provided definitive proof of its binary nature, with
the measurement of a $321\pm5\U{day}$ periodicity in the lightcurve
\cite{2011ApJ...737L..11B}. The TeV light curve displays a broad
gamma-ray flare close to the X-ray maximum, with a duration of
$\sim40\U{days}$ \cite{2011arXiv1111.2155M}. No GeV source has been detected.

A number of competing processes likely contribute to the variability
observed in the gamma-ray binaries. In particular, the efficiency of
inverse Compton gamma-ray production, as well that of VHE gamma-ray
absorption (through pair production), changes as a function of orbital
phase. This can go some way towards explaining the apparent phase
shift between the GeV and TeV lightcurves, for example in the case of
LS~5039. There are clearly other effects which contribute, however, as
demonstrated by the long term instability of
LS~I~+61${^\circ}$303. The sharp GeV cut-offs in these systems are also
difficult to explain, and may indicate that the GeV and TeV emission
components do not have the same origin.

A final comment should be reserved for the \textit{Fermi}-LAT source
1~FGL~J1018.6-5856, which was recently identified as a GeV binary
system, with a period of $16.58\pm{0.04}\U{days}$
\cite{2011ATel.3221....1C}. This source resides in a complex region,
coincident with the center of the SNR G284.3-1.8, and close to a LAT
pulsar (at a distance of 35'). The extended, unidentified HESS source,
HESS~J1018-589, overlaps with the GeV binary location, and appears to
consist of a point-like source overlaid on a diffuse structure
\cite{2010cosp...38.2803D}. While this is indicative of a new TeV
binary there is, as yet, no evidence for variability in the point-like
emission, making the identification still uncertain.

\subsection{Globular Clusters}
A single globular cluster, Terzan 5, has been suggested as the
probable counterpart of a TeV source (HESS J1747-248)
\cite{2011ICRC_Domainko}. If the association is correct, the emission
is likely related to the large population of millisecond pulsars in
this cluster, which provide an injection source of relativistic
leptons \cite{2009ApJ...696L..52V}. Inverse Compton gamma-ray photons
result when these electrons upscatter low energy photons of the
intense stellar radiation field. The H.E.S.S. source is extended, and
slightly offset from the cluster core (although there is significant
overlap). The probability of this being simply a chance positional
coincidence is $\sim10^{-4}$.

Globular clusters have also been favored targets in searches for dark
matter particle annihilation signatures, since they may have been
generated in dark matter mini-haloes before the formation of galaxies
took place, and thus retain a significant dark matter component
\cite{2009ApJS..180..330K, 2008ApJ...678..594W}. Limits have been
placed on NGC 6388, M15, Omega Centauri, 47 Tuc, M13, and M5
(\cite{2011ApJ...735...12A}, and references therein).

\subsection{Unidentified objects}

IACTs are able to locate point sources with reasonably good accuracy
(typically $\lesssim1'$ for a moderately strong source). Extragalactic
TeV sources can therefore usually be firmly identified with a single
counterpart at other wavelengths. For Galactic sources, identification
poses more of a problem, and around one third of the Galactic sources
lack a firm identification. While the diffuse Galactic background
emission, which dominates at GeV energies, is not significant, the TeV
sources themselves are mostly extended, and can often be plausibly
associated with multiple counterparts. PWN, in particular, often have
their brightest TeV emission offset from the location of the parent
pulsar or X-ray PWN, which may not yet have been detected (a number of
new energetic pulsars have been located in follow-up observations of
unidentified TeV sources). In other cases, despite deep X-ray and
radio follow-up observations, no reasonable counterpart has yet been
found. 

The unidentified Milagro sources also pose some interesting
questions. While IACTs have identified sources associated with these
objects, the TeV sources are typically much smaller in angular extent,
and cannot account for the entire Milagro flux. One possible
explanation for this is that there is a diffuse high energy component
to the emission, which is difficult to resolve with IACTs, given their
relatively small fields of view.

\section{The Future}

The field of ground-based gamma-ray astronomy has expanded
dramatically over the past 10 years, but it is worth noting that the
observatories currently operating are far from reaching the physical
limits of the detection techniques. We therefore conclude this review
with a brief discussion of some of the instrumental developments which
can be expected over the next decade.

\begin{figure}[!t]
  \vspace{0mm} \centering \includegraphics[width=\columnwidth]{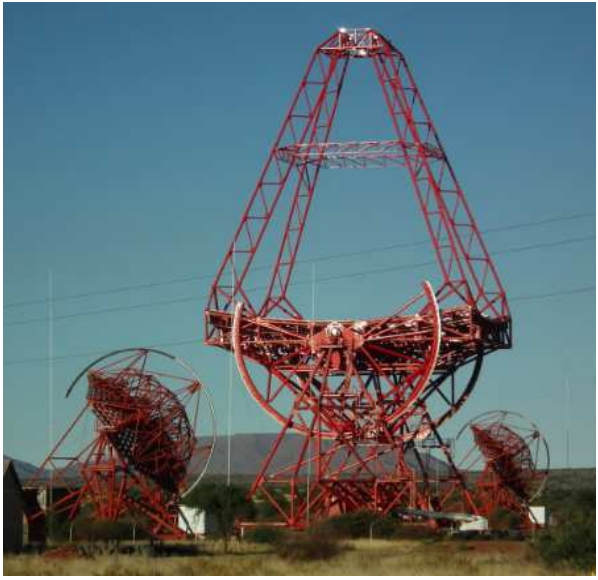}
  \caption{ The steel structure for the $600\UU{m}{2}$ H.E.S.S. II
  telescope. Mirrors and camera will be installed in the first half of
  2012. (Note that the H.E.S.S. II telescope is located in between the
  two H.E.S.S. I telescopes shown in the image, not in the
  foreground). Figure courtesy of the H.E.S.S. collaboration
  {http://www.mpi-hd.mpg.de/hfm/HESS/}}
  \label{HESS2}
 \end{figure}

The IACTs currently operating have all made significant efforts to
maintain, and improve, sensitivity since they were first
commissioned. H.E.S.S. recoated the telescope mirrors, and
successfully developed sophisticated analysis tools with which to
exploit the data. VERITAS relocated their original prototype telescope
to provide a more favorable array layout, halving the time required to
detect a weak source. Most significantly, MAGIC added a second
telescope of similar design to the first, providing a stereo pair with
a baseline of $85\U{m}$. All of these observatories have further
upgrade plans. Both MAGIC and VERITAS are implementing camera upgrades
- in the case of VERITAS this involves the replacement of all of the
camera PMTs with more sensitive, super-bialkali devices in summer
2012. H.E.S.S. are constructing H.E.S.S. II - the addition of a single
large telescope, with $600\UU{m}{2}$ mirror area, to the centre of the
array. Figure~\ref{HESS2} shows the telescope structure in November
2011. The mirrors and camera will be installed in the first half of
2012.

\begin{figure}[!t]
  \vspace{0mm} 
  \centering 
  \includegraphics[width=0.5\columnwidth]{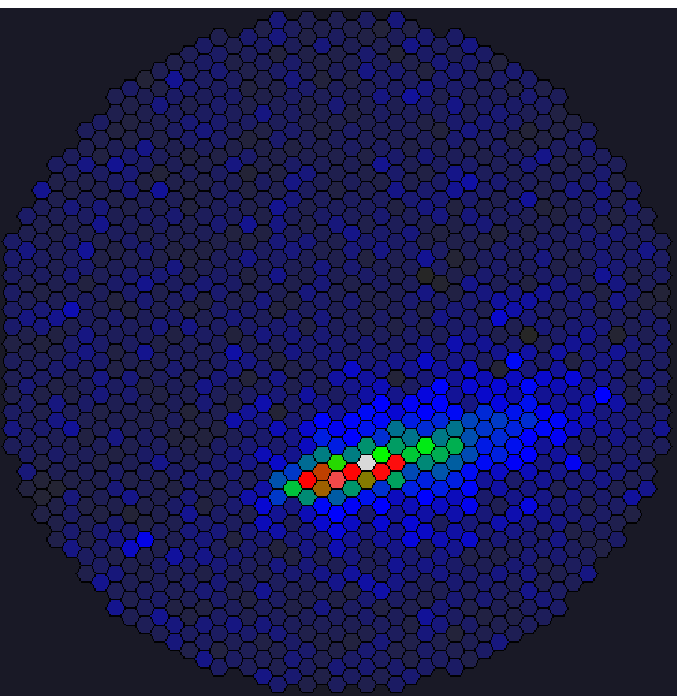}\includegraphics[width=0.503\columnwidth]{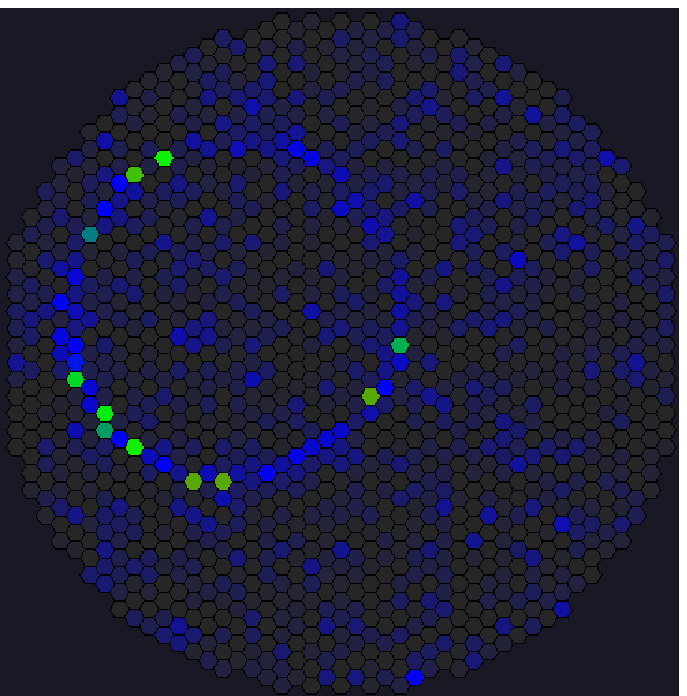}\vspace{-0.3mm} 
  \includegraphics[width=0.5\columnwidth]{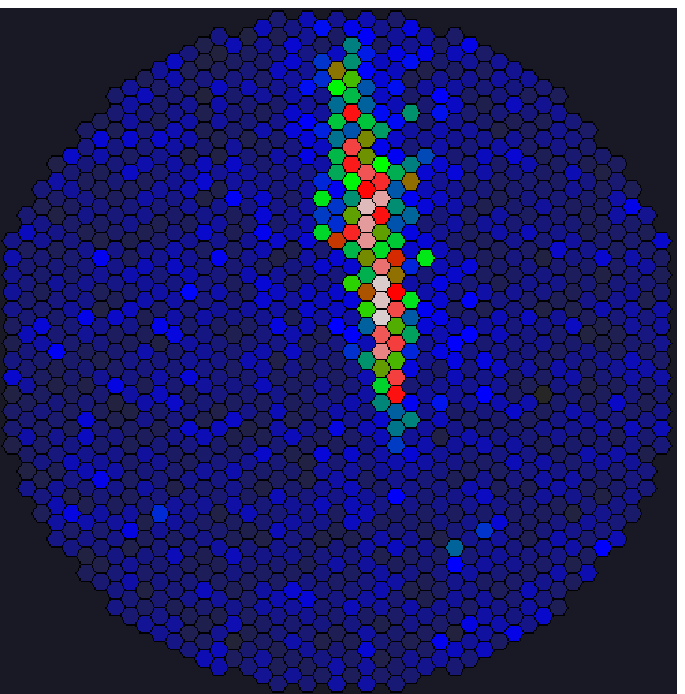}\includegraphics[width=0.5\columnwidth]{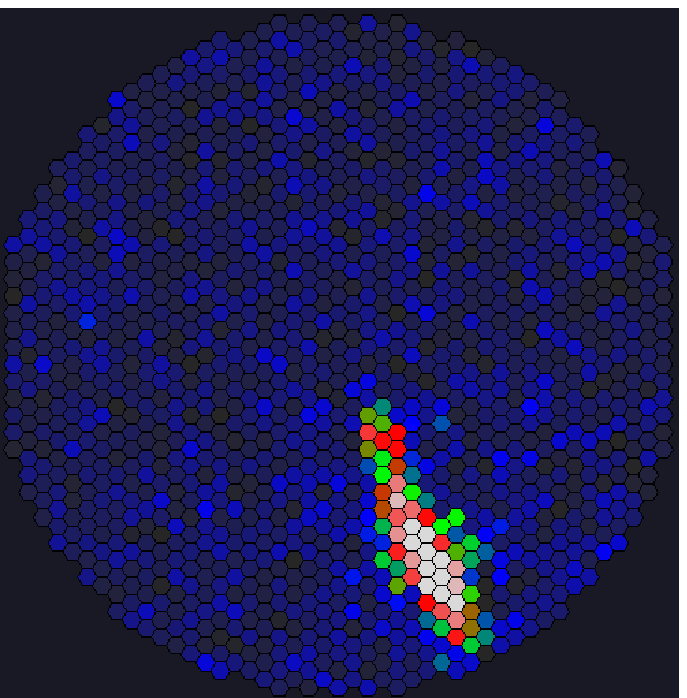}
  \caption{ First light cosmic ray images from FACT (First G-APD
  Cherenkov Telescope). Four different events are shown; the upper
  right event shows the characteristic ring image produced by a local
  muon. The camera contains 1440 Geiger-mode avalanche photodiodes,
  installed on one of the original HEGRA telescopes at the Roque de
  los Muchachos on La Palma.  Figure courtesy of the FACT
  collaboration {http://fact.ethz.ch/first}}
  \label{FACT}
 \end{figure}

Novel approaches are also being pursued by many smaller projects. A
particularly nice example of this is FACT (the First G-APD Cherenkov
Telescope), which has recently demonstrated the application of
Geiger-mode Avalanche Photodiodes for Cherenkov astronomy. The FACT
telescope consists of a 1440-pixel G-APD camera at the focus of one of
the original HEGRA telescopes. Geiger-APDs hold great promise as a
potential replacement for PMTs, since they are robust, and offer much
better photon conversion efficiency. Figure~\ref{FACT} shows some
``first light'' images from FACT. Other projects in development
include GAW (Gamma Air Watch), MACE (Major Atmospheric Cerenkov
Telescope Experiment) and LHASSO (Large High Altitude Air Shower
Observatory). LHASSO is an ambitious project which will be located
near the site of the ARGO-YBJ experiment in Tibet. It is planned to
consist of four water Cherenkov detectors, two IACTs, three
fluorescence telescopes and a large scintillator array.

For the particle detectors, the next stage in instrumentation is HAWC
(the High Altitude Water Cherenkov Observatory). HAWC will consist of
300 individual water Cherenkov tanks, at an altitude of $4100\U{m}$ in
Mexico. The final array is expected to be 15 times more sensitive than
Milagro, and will be a powerful tool for surveying, and for
observations of transient phenomena. A prototype system is already in
operation, and science operations will start in spring 2012, with
completion of the full array expected in 2014.

The most ambitious future project is CTA (the Cherenkov Telescope
Array). This is described in detail in
\cite{2011ExA...tmp..121A}. Briefly, it comprises an array of imaging
atmospheric Cherenkov telescopes covering $\sim1\UU{km}{2}$, providing
a factor of 5-10 improvement in sensitivity in the
$100\U{GeV}$-$10\U{TeV}$ range, and extending the energy range both
above and below these values. Both a northern and a southern site are
envisaged, and the array will be operated as an open
observatory. Multiple telescope designs are planned, including small
(few$\U{m}$), medium ($10-15\U{m}$) and large ($20-30\U{m}$) diameter
reflectors, as well as two-mirror telescope designs, such as the
Schwarzchild-Couder \cite{2008ICRC....3.1445V}.

In conclusion, TeV gamma-ray astronomy now describes a broad
astronomical discipline which addresses a wide, and expanding, range
of astrophysical topics. With planned instrumental developments, it is
not unreasonable to expect the source catalogue to exceed 1000 objects
within the next decade. Much of the TeV sky remains relatively
unexplored. Less than $\sim10\%$ of the sky has been observed
with $\sim10\U{milliCrab}$ sensitivity at $1\U{TeV}$, and the
sensitive exposure to transient events and widely extended sources at
these energies is much lower. The likelihood of continued exciting
results is certain, both for the known sources and source classes, and
for new discoveries.






\bibliographystyle{model1a-num-names}
\bibliography{jholder_APP_review}







\end{document}